\documentstyle[prd,eqsecnum,aps]{revtex}
\begin{document}
\title{Understanding initial data for black hole collisions}
\author{Carlos O. Lousto\thanks{Electronic Address:
lousto@mail.physics.utah.edu} and Richard H. Price}
\address{
Department of Physics, University of Utah, Salt Lake City, UT 84112}
\maketitle
\begin{abstract}
  Numerical relativity, applied to collisions of black holes, starts
  with initial data for black holes already in each other's strong
  field. For the initial data to be astrophysically meaningful, it
  must approximately represent conditions that evolved from holes
  originally at large separation. The initial hypersurface data
  typically used for computation is based on mathematically
  simplifying prescriptions, such as conformal flatness of the
  3-geometry and longitudinality of the extrinsic curvature. In the case
  of head on collisions of equal mass holes, there is evidence that
  such prescriptions work reasonably well, but it is not clear why, or
  whether this success is more generally valid. Here we study these
  questions by considering the ``particle limit'' for head on
  collisions of nonspinning holes, i.e., the limit of an extreme ratio
  of hole masses. The mass of the small hole is considered to be a
  perturbation of the Schwarzschild spacetime of the larger hole,
  Einstein's equations are linearized in this perturbation, and
  described by a single gauge invariant spacetime function $\psi$, for
  each multipole.  The resulting quadrupole equations have been solved
  by numerical evolution for collisions starting from various initial
  separations, and the evolution is studied on a sequence of
  hypersurfaces. In particular, we extract hypersurface data, that is
  $\psi$ and its time derivative, on surfaces of constant background
  Schwarzschild time.  These evolved data can then be compared with
  ``prescribed'' data, evolved data can be replaced by prescribed data
  on any hypersurface, and evolved further forward in time,
  a gauge invariant measure of deviation from conformal flatness can be
  evaluated, and other comparisons can be made.
  The main findings of this study are: (i) For holes of unequal mass
  the use of prescribed data on late hypersurfaces is not successful.
  (ii) The failure is likely due to the inability of the prescribed
  data to represent the near field of the smaller hole.
  (iii) The discrepancy in the extrinsic curvature is more important
  than in the 3-geometry. (iv) The use of the more general conformally
  flat longitudinal data does not notably improve this picture.
\end{abstract}
\pacs{04.30.-w, 04.70.Bw}

\section{Introduction and background}
The collision of black holes is of great interest as both astrophysics
and as a strong-field gravitational interaction with no Newtonian
analog.  Black hole collisions also may provide the strongest, most
observable, source of gravitational radiation that can be detected by
gravitational wave observatories now under construction\cite{ligo}.
The strong field nature of the process means that the nonlinear
character of Einstein's theory plays a crucial role, so that estimates
based on linearized theory are unreliable.  For the last stage of
coalescence of two rotating holes in a decaying orbit, we have only
dimensional estimates. The timeliness of the problem has given rise to
an effort to attack the problem by solving Einstein's equations
numerically on supercomputers\cite{grandchallenge}. 

The numerical solution starts with some ``initial value'' data, a
solution of a subset of Einstein's equations on an initial
hypersurface, and then evolves this solution forward in time.  The
specification of the initial data is tantamount to specifying what
problem it is that the computer will be solving, so the initial data
must encode the physical parameters (mass, spin, location, momentum)
of the colliding objects. These physical parameters are meaningful and
well understood for isolated holes, but become increasingly
ambiguous as the initial separation between the holes decreases, and
the field of each hole has a strong effect on the spacetime geometry
of the other.  Ultimately, at sufficiently small initial separation,
it becomes impossible to make any clear statements about the physical
parameters of an individual hole.

In principle, one would like to avoid this difficulty by specifying
initial data on a hypersurface in the distant past when the holes were
separated by a distance many times the horizon size. When the initial
influence of each hole on the other becomes small, the
uncertainties in the choice of initial data become negligible.
Unfortunately, it is not possible to start the numerical evolution a
long time before the collision. Einstein's equations are a nonlinear
hyperbolic system, and for such systems instabilities in numerical
evolution seem to be a generic feature.  While it can be expected that
some progress will be made in improving the codes and suppressing the
instabilities, it is not likely that in the near future these codes
will be able to evolve initial data for more than a few dynamical
times, the characteristic time scale for black hole processes (around
$10^{-5}$\,sec for a solar mass hole).  This limitation precludes
starting the dynamics with the holes at large distances and means that
the initial data must be given on a hypersurface on which the
hole-hole interactions are already strong. If the resulting numerical
evolutions are to be astrophysically meaningful, it is crucial to have
a way of connecting data on this ``late'' hypersurface to an astrophysical
precursor configuration with the holes interacting weakly. In other words,
to understand the meaning of initial data on a ``late'' hypersurface, we
must know where that data came from.
One approach to providing this connection is to use approximations
such as higher-order post-Newtonian methods in order to understand the
evolution of the system from an early stage to a stage at which the
holes are 
interacting with intermediate strength.  Such an approach may turn
out to be sufficient if the numerical evolutions can be stabilized for
a relatively long time.  

There is reason to hope that it might be possible to give initial data
at late times, with relatively little difficulty. The reason is based
on the success of a ``two phase'' approximation method used by
Abrahams and Cook\cite{abcook} and further investigated by Baker and
Li\cite{bakerli}.  These studies dealt with the head on (zero angular
momentum) collisions of two equal-mass nonrotating holes. Data were
specified on a very late hypersurface, simply by using a ``standard''
initial value solution formally representing equal mass holes moving
towards each other. This standard solution contained a parameter $P$
which, in the case of an isolated hole, agreed with the momentum of
the hole. On the very late hypersurface the separation and momentum
were set at the values dictated by Newtonian gravity theory if the
holes had started infall from some large initial separation. The
standard initial value data was evolved forward in time (with the use
of an approximation method). The radiated energy found by this
approximation could be compared with the ``correct'' radiated energy,
since numerical relativity results were available for such collisions.
The approximation method agreed with numerical relativity to rather
good accuracy. This is somewhat a surprise since the ``standard''
numerical relativity initial data should be very different from the
``correct'' evolved data.  The implication of this agreement is that
results for gravitational radiation might not be sensitive to all details
of initial data.

We explore this question in the present paper by comparing
prescriptions for imposing data on hypersurfaces with the ``correct''
data evolved from an earlier configuration. In order to do this, of
course, we need the ability to evolve a solution for a relatively long
time, something that cannot yet be done with the nonlinear Einstein
equations.  We choose therefore to use the particle approximation, the
approximation in which the mass $m$ one of our holes, is much smaller
than the mass of the other $M$. We then treat the low-mass hole as a
particle of negligible size, and we treat $m/M$ as an expansion
parameter. By doing first-order perturbation theory with this
parameter, our evolution equations become linear, so that a stable
numerical scheme is straightforward to develop.

In an earlier paper\cite{paperI} (Paper I), we developed the basic
mathematics of the particle limit approximation, and of evolving
data that was initially stationary. In that paper, Laplace transform
methods were used. The method for the present results is a direct
numerical solution of the partial differential equations in radius and
time (after multipole decomposition). We have compared results of the
two completely different methods, and found agreement within the
estimated numerical accuracy, thereby adding confidence that there are
no mistakes in our numerical methods.

The paper is organized as follows. In Sec.\ II we give the
mathematical formulation of the problem.  We base this formulation on
the gauge-invariant Moncrief\cite{moncrief} approach to perturbations,
so that the variable $\psi$ we use is invariant with respect to
perturbative changes in slicing. We need only specify the slicing to
zero order in the perturbation, and to that order we take our
hypersurfaces to be slices of constant Schwarzschild time. In Sec.\ II
we discuss in detail the freedom in initial data and the standard
choices made for reducing that freedom.  We go on to explain the
particular physically-motivated choice that we make for arriving at a
definitive prescription of initial data. In Sec.\ III, we start by
explaining our numerical method for evolving $\psi$.  The scheme for
solving the linear hyperbolic equation obeyed by $\psi$ is
straightforward. The only issue that deserves attention is the source
term in the equation for $\psi$. In the particle limit, the small hole
is treated as a point particle, and this particle is associated with a
stress energy source that enters the $\psi$ equation as the
derivative of a delta function of radial position. We describe in
Sec.\ III how the singular source is handled. Numerical results are
presented and compared in Sec.\ III.B. In Sec.\ IV we discuss the
results and their meaning.

\section{Mathematical formulation}
\subsection{Moncrief-Zerilli formalism}

We will describe the perturbations to the Schwarzschild background,
due to the particle, in the notation of Regge and Wheeler\cite{rw}
(hereafter RW).  The symmetry of the straightline infall of the
particle means that there will be no odd parity perturbations. For the
even parity perturbations 
the general form of the line element, with perturbations of a specific
multipole index $\ell$, is
\begin{eqnarray}\label{rwform}
ds^2=ds^2_0+(1-2M/r)(H_0^\ell Y_{\ell 0})dt^2+ (1-2M/r)^{-1}(H_2^\ell
Y_{\ell 0})dr^2\nonumber\\ +r^2(K^\ell Y_{\ell 0}+G^\ell
\partial^2Y_{\ell 0}/\partial\theta^2)d\theta^2 +r^2(\sin^2\theta
K^\ell Y_{\ell 0}+G^\ell \sin\theta\cos\theta\partial Y_{\ell
0}/\partial\theta )d\phi^2\nonumber\\ +2H_1^\ell Y_{\ell 0}dtdr
+2h_0^\ell (\partial Y_{\ell 0}/\partial\theta) dtd\theta +2h_1^\ell
(\partial Y_{\ell 0}/\partial\theta) drd\theta\ .
\end{eqnarray}
Here $ds^2_0$ is the unperturbed line element for a Schwarzschild
spacetime of mass $M$, where $H_0^\ell ,H_1^\ell ,H_2^\ell ,h_0^\ell
,h_1^\ell ,K^\ell $ and $G^\ell $ are functions of $r,t$, and $Y_{\ell
0}(\theta)$ are the $m=0$ spherical harmonics. For simplicity, from
here on we shall drop the $\ell$ index on perturbation functions.

There are two closely related formalisms available for describing the
evolution of the perturbations in terms of a single wave function and
a single wave equation. The method due to Moncrief\cite{moncrief} uses
only information ($H_2,h_1,K,G$) about the 3-geometry of a
$t$=\,constant hypersurface, and is gauge invariant,
i.e., independent of the choice of lapse and shift, and of
diffeomorphisms on the hypersurface. The second method, due to
Zerilli\cite{zerilli}, relies on a specific gauge, the gauge choice
introduced by Regge and Wheeler\cite{rw}, in which $h_0,h_1$, and $G$
are set to zero.

As in Paper I we shall use the Moncrief formalism.
The Moncrief wave function $\psi$, in terms of the perturbations in the RW 
notation, is
\begin{equation}\label{psidef}
  \psi(r,t)=\frac{r}{{\lambda}+1}\left[
    K+\frac{r-2M}{{\lambda}r+3M}\left\{ H_2-r\partial K/\partial r
    \right\} \right]
+2(1-2M/r)\left(r^2\partial G/\partial r-2h_1
\right)\ ,
\end{equation}
where we have used Zerilli's normalization for $\psi$ and his notation
\begin{equation}{\lambda}\equiv(\ell+2)(\ell-1)/2\ .
\end{equation}
For simplicity, we shall occasionally present some equations restricted to the
RW gauge. Since the Moncrief wave function is gauge invariant, this gauge
restriction allows simplicity of presentation, with no other
consequences. In this gauge, we omit the $G$ and $h_1$ terms in 
Eq.\ (\ref{psidef}).

The basic wave equation for an infalling particle is given in Paper I
as 
\begin{equation}\label{rtzerilli}
-\frac{\partial^2\psi}{\partial t^2}
+\frac{\partial^2\psi}{\partial r*^2}-V_{\ell}(r)\psi=
{\cal S}_\ell (r,t)\ .
\end{equation}
Here $r^*\equiv r+2M\ln(r/2M-1)$ is the Regge-Wheeler\cite{rw}
``tortoise'' coordinate and $V_\ell$ is the Zerilli potential (given,
e.g., in Paper I). For a point particle of proper mass $m_0$, the
stress energy is given by
\begin{equation}\label{tmunu}
T^{\mu\nu}=(m_0/U^0)U^\mu U^\nu \delta(r-r_p[t])\delta^2(\Omega)/r^2\ ,
\end{equation}
where $U^\mu$ is the particle 4-velocity.
The two dimensional delta function
$\delta^2(\Omega)$ gives the angular location of the particle
trajectory
\begin{equation}
\delta^2(\Omega)=\sum_{\ell,m}Y_{\ell m}(\theta,\phi)Y_{\ell m}^*
(\theta,\phi)=\sum_\ell Y_{\ell0}(\theta)\sqrt{(2\ell+1)/4\pi}\ ,
\end{equation}
with the last expression applying for infall along the positive $z$
axis.  The time dependent location of the particle $r_p(t)$ follows from
the geodesic equation and, for a particle starting from rest at $t=0,
r=r_0$, is the inverse of
\begin{eqnarray}\label{time}
T(r_p)&=& \sqrt{1-\frac{2M}{r_0}}\left({r_0\over2M}\right)
\left({r_p\over2M}\right)^{1/2}
\sqrt{1-{r_p\over r_0}}
\nonumber\\
&&+(1+{4M\over r_0})\left({r_0\over2M}\right)^{3/2}\sqrt{1-\frac{2M}{r_0}}
\arctan\left[\sqrt{{r_0\over r_p}-1}\right]\nonumber\\
&& \ln{\left[
\frac{1-\left(1-4M/r_0\right)(r_p/2M)
+2\sqrt{1-2M/r_0}\sqrt{r_p/2M}
\sqrt{1-r_p/r_0}}{(r_p/2M)-1}
\right]}\ .
\end{eqnarray}

{}From the particle stress energy in (\ref{tmunu}), the source term
${\cal S}_\ell (r,t)$ on the right hand side of (\ref{rtzerilli}), is
given in Paper I as
\begin{eqnarray}\label{rtsource}
{\cal S}_\ell (r,t)=-\frac{2(1-2M/r)\kappa
}{r(\lambda+1)(\lambda r+3M)
}\left[-r^2(1-2M/r)\frac{1}{2 U^0}\delta'(r-r_p[t])\right.\nonumber\\
+\left.\left\{\frac{r(\lambda+1)-M}{2 U^0}-\frac{3M U^0 r(1-2M/r)^2
}{\lambda r+3M}\right\}\delta(r-r_p[t])
\right]\ ,
\end{eqnarray}
where
\begin{equation}
\kappa\equiv8\pi m_0\sqrt{(2\ell+1)/4\pi}\ .
\end{equation}

The total radiated energy after a given $t_0$ is
\begin{equation}\label{wenergy}
  {\rm Energy}=\frac{1}{64\pi}\frac{ (\ell+2)!
    }{(\ell-2)!}\int_{t_{0}}^\infty
  \left(\dot\psi\right)^2\,dt\ .
\end{equation}

\subsection{Hypersurface data for $\psi$}

We choose our data, on a hypersurface of constant $t$, to
correspond to the choice usually made for work in numerical
relativity. The data, for both the initial 3-geometry and extrinsic
curvature, are based on the work of York\cite{york} and collaborators.
This approach takes the 3-geometry of the initial hypersurface to be
given by $d\sigma^2=\Phi^4d\sigma^2_{\rm flat}$. The conformal factor
$\Phi$ must, in vacuum, satisfy the equation
\begin{equation}\label{genham}
\nabla^2\Phi=-\frac{1}{8}\Phi^{5}K_{ij}K^{ij}\ ,
\end{equation}
where $\nabla^2$ is the Laplacian taken with respect to the flat
background, and where $K_{ij}$ is the extrinsic curvature. 

For a constant $t$ slice of the unperturbed Schwarzschild spacetime,
the extrinsic curvature vanishes. It follows that $K_{ij}$ is
perturbative, and hence the right hand side of (\ref{genham}) is a
higher order perturbation, which, in our first order perturbation
calculation, we can ignore.  We denote coordinates in the conformally
related flat space with bars, such as $\bar{r}, \bar{z}$, and we note
that $\Phi$ must be a harmonic function of these coordinates.  We make
two different choices for $\Phi$. The first is the solution discussed
by Brill and Lindquist\cite{bl}, and which we will call a BL type
solution. This solution follows if we treat $2\Phi$ as if it were the
Newtonian potential of two points, one of mass $M$, and one of mass
$m_p$. If the first mass is located at the origin of the flat
coordinates, and the second is at (flat)
coordinate position $\bar{z}=\bar{z}_p$, then the BL solution takes the form
\begin{equation}\label{BLPhi}
\Phi_{\rm BL}=1+\frac{M}{2\bar{r}}+
\frac{m_p/2}{|\bar{r}{\bf e}_r-\bar{z}_p{\bf e}_z|}\ .
\end{equation}
Though the origin and the point $\bar{r}{\bf e}_r=\bar{z}_p{\bf e}_z$
are coordinate singularities, they are not geometric singularities.
Near these coordinate points the divergence of the conformal factor
means that the ``points'' are, in fact, asymptotically flat regions,
so the geometry given by (\ref{BLPhi}) is actually that of three
asymptotically flat spaces connected by two throats.  For our purposes
here, we take the ``particle limit,'' the limit that $m_p\ll M$.

The BL solution is the only choice that was considered in Paper I.  A
second initial conformal factor that can be considered is that which
is ``reflection symmetric,'' in the sense that there are only two
asymptotically flat regions, and they are isometric to each other. The
method of constructing such solutions was given by
Misner\cite{misnergen}, and requires that ``image'' points be used in
the conformal space, in such a way that the geometry is symmetric with
respect to inversion through a sphere about each of the point
singularities.  Here we make the solution symmetric only for a sphere
about the origin. The singularity at $\bar{r}{\bf e}_r=\bar{z}_p{\bf
  e}_z$, in the limit, will be taken to have a zero size inversion
sphere around it (the particle limit) and the images inside it will be
meaningless. (Alternatively one can construct a solution reflection
symmetric for two spheres, and subsequently take the limit; the result
is the same.) The conformal factor, for this ``Misner'' case, is
\begin{equation}\label{MisPhi}
\Phi_{\rm Mis}=1+\frac{M}{2\bar{r}}+
\frac{m_p/2}{|\bar{r}{\bf e}_r-\bar{z}_p{\bf e}_z|}
+\frac{m_{\rm image}/2}{|\bar{r}{\bf e}_r-\bar{z}_{\rm image}{\bf e}_z|}\ .
\end{equation}
It is straightforward to verify that this Misner solution is symmetric
for inversion about ${\bar r}=M/2$ if the parameters $m_{\rm
  image},\bar{z}_{\rm image}$ are chosen to be $m_{\rm
  image}=m_pM/(2\bar{z}_p)$ and $\bar{z}_{\rm
  image}=(M/2)^2/\bar{z}_p$.

For either the BL form or the Misner form the line element for the 3-geometry
can now be written
\begin{equation}\label{rbarform}
d\sigma^2=\left[1+M/2\bar{r}+\sum_{\ell=0,1,2,\cdots}\alpha_\ell(\bar{r})
P_\ell(\cos{\theta})\right]^4\left(d\bar{r}^2+\bar{r}^2d\Omega^2\right)\ ,
\end{equation}
with $d\Omega^2=d\theta^2+\sin^2\theta d\phi^2.$ The requirement that
$\Phi$ be harmonic means that $\alpha_\ell$ must have the form $a_\ell
{\bar r}^\ell+b_\ell {\bar r}^{-(\ell+1)}$.  We must now put the
metric into a form for comparison with the Schwarzschild metric. 
To do this we introduce a Schwarzschild-like radial coordinate $r$
related to the flat space coordinate $\bar{r}$ by
\begin{equation}\label{rvsrbar}
\bar{r}=\left(\sqrt{r}+\sqrt{r-2M}
\right)^2/4\,,\hspace*{35pt} r=\bar{r}\left(1+M/2\bar{r}
\right)^2\ .
\end{equation}
This is only
one of many possibilities for such a transformation, but it is the
simplest for computation, and has been very convenient for
perturbation analysis\cite{pp}. This transformation puts
(\ref{rbarform})
into the form
\begin{equation}\label{gends2}
d\sigma^2=\left[1+\frac{2m_p/{\bar r}}{1+M/(2\bar{r})}\sum_{\ell=0, 1,2\cdots}
{\cal F}_\ell(\bar{r})P_\ell(\cos{\theta})
\right]\left(\frac{dr^2}{1-2M/r}+r^2d\Omega^2
\right)\ ,
\end{equation}
where we have kept only terms first order in $m_p$, and, as in Paper
I, have introduced the notation  ${\cal F}_\ell({\bar
r})= (2{\bar r}/m_p)\alpha_\ell({\bar r})$.

The perturbations, in the RW notation, can now be written
\begin{equation}\label{initK}
K=H_2=\frac{2m_p/\bar r}{1+M/2\bar r}
{\cal F}_\ell(\bar r)
\sqrt{\frac{4\pi}{2\ell+1}}
\ ,
\end{equation}
with $h_1=G=0$. These perturbations of the initial hypersurface turn out
to be in the RW gauge, so we can compare them with the hamiltonian
constraint, in the RW gauge, as given by Zerilli\cite{zerilli2},
and in Paper I, for a particle of proper mass $m_0$.
\begin{eqnarray}\label{constraint}
\left(
1-\frac{2M}{r}
\right) \frac{\partial^2K}{\partial r^2}
+
\left(
3-\frac{5M}{r}
\right) \frac{1}{r}
\frac{\partial K}{\partial r}
- \frac{\ell(\ell+1)}{2r^2}\left(K+H_2\right)
-\frac{1}{r^2}\left(H_2-K\right)
\nonumber\\ 
-\left(1-\frac{2M}{r}\right)
\frac{1}{r}\frac{\partial H_2}{\partial r}
=-8\pi m_0\sqrt{\frac{2\ell+1}{4\pi}} U^0\left(
1-\frac{2M}{r}
\right)\frac{1}{r^2}\delta(r-r_p)\ ,
\end{eqnarray}
in which the right hand side comes directly from the stress energy
expression (\ref{tmunu}). 

For the choice of a conformally flat 3-geometry 
we have $H_2=K$ and Eq.\,(\ref{constraint}) becomes a differential
equation for $K$.  A particular solution of this equation, well
behaved at infinity, is
\begin{equation}\label{particular}
K^{\rm partic}=\sqrt{\frac{4\pi}{2\ell+1}}\,
\left(\frac{2m_p/\bar r}{1+M/2{\bar r}}\right)
\left\{
\begin{array}{ll}
\left(\bar{z}_p/{\bar r}\right)^\ell&\ \ {\bar r}>\bar{z}_p\\ 
\left({\bar r}/\bar{z}_p\right)^{(\ell+1)}&\ \
{\bar r}<\bar{z}_p
\end{array}
\right.\ .
\end{equation}
This solution, in fact, corresponds precisely the BL solution of Eq.\ 
(\ref{BLPhi}).  For $\bar{z}_{\rm image}<M/2$ and $\bar{r}>M/2$, a
homogeneous solution well behaved at infinity is
\begin{equation}\label{homog}
K^{\rm homog}=\sqrt{\frac{4\pi}{2\ell+1}}\,
\left(\frac{2m_{\rm image}/\bar{r}}{1+M/2{\bar r}}\right)
\left(
\frac{\bar{z}_{\rm image}}{\bar r}
\right)^\ell\ .
\end{equation}
If we take $\bar{z}_{\rm image}=(M/2)^2/\bar{z}_p$ and $m_{\rm
  image}=m_pM/(2\bar{z}_p)$, then the homogeneous solution can be
considered to be the solution due to an ``image'' mass, inside the
horizon at $\bar{r}=\bar{z}_{\rm image}$, and it is straightforward to
verify that with these choices of $\bar{z}_{\rm image}$ and $m_{\rm
  image}$ the Misner solution is $K^{\rm Mis}=K^{\rm partic}+K^{\rm
  homog}$. There are, of course, other possibilities. We could, for
instance, consider, with the same choices of $\bar{z}_{\rm image}$ and
$m_{\rm image}$, the combination $K^{\rm anti} =K^{\rm partic}-K^{\rm
  homog}$. For this solution $K$ and $\psi$ would vanish on the
horizon. This property turns out to be preserved by evolution as is
made clear in the discussion below.

By using (\ref{rvsrbar}) and integrating across the singularity in
(\ref{constraint}) at $r=r_p =\bar{z}_p[1+M/(2\bar{z}_p)]^2$, we find that
\begin{eqnarray}\label{jumpcond}
\Delta K,_r\equiv dK/dr|_{r=r_p^+}-dK/dr|_{r=r_p^-}\nonumber\\
=-8\pi m_p\frac{\sqrt{(2\ell+1)/4\pi}} {\bar{z}_p^{1/2}
r_p^{3/2}\sqrt{1-2M/r_p}}
\\ =-8\pi m_0\nonumber
\frac{\sqrt{(2\ell+1)/4\pi}}{r_p^2}U^0\ .
\end{eqnarray}
The 4-velocity component $U^0$
is given by
\begin{equation}
U^0=\frac{\sqrt{1-2M/r_0}}{1-2M/r_p}\ ,
\end{equation}
so it follows from 
(\ref{jumpcond}) that
\begin{equation}\label{finalm2m0}
m_p=m_0\sqrt{
\left(\frac{\bar{z}_p}{r_p}\right)
\frac{1-2M/r_0}{1-2M/r_p}
}={\textstyle\frac{1}{2}}m_0
\left(1+\sqrt{1-\frac{2M}{r_p}}
\right)
\sqrt{
\frac{1-2M/r_0}{1-2M/r_p}}\ .
\end{equation}
Note that the relationship of the mass parameter $m_p$ and the particle
mass $m_0$ is the same for the Misner case as for the BL case, and hence 
the same as in Paper I. This must be true, of course, since the 
symmetrizing image term addition in 
(\ref{homog}) has no discontinuity at $r=r_p$. 

The initial hypersurface, on which the particle and the hypersurface
data are stationary, is always denoted by $t=0$.
On hypersurfaces with $t>0$ we shall limit our choice
of prescribed data to conformally
flat 3-geometries. The choice of conformal flatness is not, however,
preserved by evolution. This means that the 3-geometry on 
constant $t$ hypersurfaces
will in general not be conformally flat. On such a 
hypersurface, we will have numerical values only for a single
function, the Moncrief function $\psi$. It turns out that we
can test $\psi$ for underlying conformal flatness. The combination of
metric perturbations
\begin{equation}\label{confindex}
I_{\rm conf}\equiv
H_2-K+\frac{2}{r}\left(1-\frac{3M}{r}
\right)\ \left(
h_1-\frac{r^2}{2}\partial_rG
\right)
-2\left(1-\frac{2M}{r}
\right)\partial_r\left(
h_1-\frac{r^2}{2}\partial_rG
\right)
\ .
\end{equation} 
is gauge invariant, and clearly vanishes for a 3-geometry that is in
conformally flat form, with $h_1=0, G=0$ and $H_2=K$. [See Eq.\ 
(\ref{rwform}).]  The computation of this gauge invariant
quantity from $\psi(r)$ is most easily described in the RW gauge, where
$I_{\rm conf}$ reduces to $H_2-K$. From Eq.\ (\ref{psidef}) in the RW gauge,
and from Eq.\ (\ref{constraint}), which is already in the RW gauge, it
follows that (for $r\not=r_p)$
\begin{equation}\label{H2Kextract}
K=\frac{
6M^2+3M\lambda r+\lambda(\lambda+1)r^2
}{r^2(\lambda r+3M)}\psi
+\left(
1-\frac{2M}{r}
\right)\,\frac{\partial\psi}{\partial r}
\ .
\end{equation} From this result for $K$, and from Eq.\ (\ref{psidef})
in the RW gauge, one finds $H_2$ to be (for $r\not=r_p)$
\begin{eqnarray}\label{H2extract}
H_2=-\frac{9M^3+9\lambda M^2r+
3\lambda^2Mr^2+\lambda^2(\lambda+1)r^3}{r^2(\lambda r+3M)^2}\, \psi
+\frac{3M^2-\lambda Mr+\lambda r^2}{r(\lambda r+3M)}\partial_r\psi 
+(r-2M)\partial^2_r\psi
\ .
\end{eqnarray} 
The difference between the expressions in Eqs.\ (\ref{H2extract}) and
(\ref{H2Kextract}) gives $I_{\rm conf}$, the gauge invariant measure of
the deviation of the 3-geometry from conformal flatness.

\subsection{Hypersurface data for  $\dot{\psi}$}
 
The integration of the wave equation (\ref{rtzerilli}), to evolve
forward in time from a $t=$\, constant hypersurface, requires that we
specify both $\psi$ and $\dot{\psi}$ at the hypersurface. In Paper I
our initial hypersurface was one of time symmetry, so we had
$\dot{\psi}=0$ initially. We now take the particles to be moving on
the hypersurface, and for the hypersurface value of the extrinsic
curvature we follow the prescription of Bowen and
York\cite{bowenyork}.  In that prescription one chooses a slicing such
that the trace $K^i_i$ of the extrinsic curvature vanishes, and one
defines a quantity $\widehat{K}_{ij}$, related to the extrinsic
curvature $K_{ij}$ by
\begin{equation}\label{KvsKhat}
K_{ij}=\Phi^{-2}\widehat{K}_{ij}
\ ,
\end{equation}
where $\gamma_{ij}^{\rm flat}$ is the flat 3-metric.  The curvature
measure $\widehat{K}_{ij}$ is considered to be a tensor in the
conformally related flat space; its indices are raised and lowered
with $\gamma_{ij}^{\rm flat}$.   The
momentum constraint turns out to be equivalent to the
requirement that $\widehat{\nabla}_i\widehat{K}^{ij}$ vanish outside
the sources, which we take to be points in the conformally flat
3-geometry. The Bowen-York prescription for initial data is
that $\widehat{K}_{ij}$ be ``longitudinal,'' i.e., derivable from a
vector $\widehat{W}_{i}$ according to
\begin{equation}\label{KfromW}
\widehat{K}_{ij}=\widehat{\nabla}_i\widehat{W}_j
+\widehat{\nabla}_j\widehat{W}_i-(2/3)\gamma^{\rm
flat}_{ij}\widehat{\nabla}_k\widehat{W}^k\ .
\end{equation}

Our choice to describe a moving hole is a solution closely related to
those used in most numerical relativity simulations. To describe these
solutions we use the geometry of the conformally related flat space
pictured in Fig.~\ref{fig:geomforK}.  The symmetry axis is taken to be
the $\bar{z}$ coordinate axis and the perturbative particle is located at
$\bar{z}=\bar{z}_p$. The vectors $\vec{\bar{r}}$ and $\vec{\bar{\rho}}$ are
the displacements (in the flat space) to a field point,
respectively from the coordinate origin and from the location of the
moving ``particle.'' The unit vector $\vec{n}$, defined
to be $\vec{\bar{\rho}}/|\vec{\bar{\rho}}|$ is the direction to the
field point from the moving particle.  In terms of the notation in this figure,
we define two solutions of the momentum constraint
$\widehat{\nabla}_i\widehat{K}^{ij}=0$. The first, which we shall call
the basic solution, $\widehat{K}^{\rm basic}$, is
\begin{equation}\label{Kfund}
\widehat{K}^{\rm basic}_{ij}=\frac{3}{2\bar{\rho}^2}\left[
P_i n_j+P_j n_i-\left(\gamma^{\rm flat}_{ij}-n_i n_j
\right)P^k n_k
\right]\ .
\end{equation}
This is a longitudinal solution that corresponds to the vector
\begin{equation}\label{Wfund}
\widehat{W}^{\rm basic}_i=
-\frac{1}{4\bar{\rho}}\left[
7P_i+n_iP^k n_k
\right]\ .
\end{equation}
The parameter $\vec{P}$ is
the momentum of a moving hole in the conformally related flat space.
For our particle, $\vec{P}$ is related to the true 3-momentum (i.e.,
the spatial part of the 4-momentum) $\vec{p}$ by $|\vec{P}|=\Phi^2|\vec{p}|$.
Here $\Phi$ is the conformal factor and, to lowest order, has the form
$1+M/2\bar{r}$, as in Eq.\ (\ref{gends2}).  In our head on collision
the only nonvanishing component of momentum is the radial component,
so we have
\begin{equation}\label{Pvspr}
P\equiv|\vec{P}|=\Phi^2p^r/\sqrt{1-2M/r},
\end{equation}
with $p^r\equiv m_0U^r$ the usual contravariant radial component of
the 4-momentum in the Schwarzschild background coordinates. This
relationship can be verified by using the extrinsic curvature given by
Eq.\ (\ref{Kfund}) in the momentum constraint $R_{tr}=8\pi T_{tr}$ and
taking the right hand side to be the stress energy of a radially
moving particle.

The ``basic'' solution in Eq.\ (\ref{Kfund}) is the simplest solution
of the momentum constraints. It is straightforward, though tedious, to
decompose the conformal extrinsic curvature into spherical harmonics.
To express the results we adopt the following notation, analogous to
that of Regge and Wheeler\cite{rw} for the metric perturbations:
\begin{eqnarray}\label{Knotation}
\widehat{K}_{\bar{r}\bar{r}}&=&\sum_{\ell=1,2,\cdots}
{\cal A}_\ell(\bar{r};\bar{z}_p)Y_{\ell0}\\
\widehat{K}_{\bar{r}\theta}&=&\sum_{\ell=1,2,\cdots}
{\cal B}_\ell(\bar{r};\bar{z}_p)\partial 
Y_{\ell0}/\partial\theta\\
\widehat{K}_{\theta\theta}&=&\bar{r}^2\sum_{\ell=1,2,\cdots}
\left[ {\cal K}_\ell(\bar{r};\bar{z}_p)Y_{\ell0}
+{\cal G}_\ell(\bar{r};\bar{z}_p)\partial 
Y_{\ell0}/\partial^2\theta^2\right]
\\
\widehat{K}_{\phi\phi}&=&\bar{r}^2\sin^2\theta\sum_{\ell=1,2,\cdots}
\left[ {\cal K}_\ell(\bar{r};\bar{z}_p) Y_{\ell0}
+{\rm cot}\theta\,{\cal G}_\ell(\bar{r};\bar{z}_p)\partial
Y_{\ell0}/\partial\theta\right]\ .\label{Knot_last}
\end{eqnarray}
Other components vanish due to the azimuthal symmetry.

For the expansion of the basic extrinsic curvature, the explicit
expressions for the coefficient functions are
\begin{eqnarray}\label{fundCals}
{\cal A}^{\rm basic}_\ell&=&\sqrt{\frac{ 4\pi }{2\ell+1}}
\left(\frac{P}{2\bar{r}^2}
\right)q^{l-1}(l+1)(l+2) 
\left[-{l\over2l-1}+q^2{(l-6)\over2l+3}
\right]\\
{\cal B}^{\rm basic}_\ell&=&\sqrt{\frac{ 4\pi }{2\ell+1}}
\left(\frac{P}{2\bar{r}}
\right)q^{l-1}(l+2) 
\left[{(l-2)\over2l-1}-q^2{(l-6)\over2l+3}
\right]\\
{\cal G}^{\rm basic}_\ell&=&\sqrt{\frac{ 4\pi }{2\ell+1}}
\left(\frac{P}{2\bar{r}^2}
\right)q^{l-1}(l+1)
\left[{5l\over2l-1}-q^2{(l-6)\over2l+3}
\right]\\
{\cal K}^{\rm basic}_\ell&=&\sqrt{\frac{ 4\pi }{2\ell+1}}
\left(\frac{P}{2\bar{r}^2}
\right)q^{l-1}
\left[-{(l-8)\over2l-1}+q^2{(l-6)\over2l+3}
\right]\label{lastbasic}\ .
\end{eqnarray}
Here $q\equiv \bar{z}_p/\bar{r}<1$ and
for $q>1$, the above expressions are valid if $\ell$ is replaced
throughout
by $-\ell-1$ everywhere on the righthand side, except in the normalization
constant $\sqrt{4\pi/2\ell+1}$.

The parameter $P$ is taken to represent the (perturbative) momentum of
the particle.  We take our asymptotically flat coordinate system to be
one in which there is no net momentum. This means that the large
``background'' hole must have momentum which is in some sense equal in
magnitude to the momentum of the particle. If we take the origin of
our coordinate system, in the conformally flat space, to be (in some
sense) at the center of mass, then the coordinate singularity
representing the background hole is at a coordinate distance
$\bar{z}_{bh}=-(m_p/M)\bar{z_p}$ from the origin. We can get the
contribution to the extrinsic curvature from the background hole by
reversing the sign of $P$ in the expressions above, and by
substituting $\bar{z}_{bh}$ in place of $\bar{z}_p$. But $z_{bh}$ is a
first order quantity, so $(\bar{z}_{bh})^n$ terms can be kept only
when $n=0$. From Eq.\ (\ref{fundCals})-(\ref{lastbasic}) we see that
this means that only $\ell=1$ terms can be kept. But the $\ell=1$ even
parity terms have no physical content, and are not coupled to
gravitational radiation. In treating the perturbative extrinsic
curvature, therefore, we consider only the contributions from the
particle.

To use the above results in calculations, we must relate the extrinsic
curvature to the Moncrief wave function $\psi$. It is useful to have
at hand the relations between the extrinsic curvature and the metric
perturbations. This is found by considering a $t=$\,constant slice of
the perturbed spacetime. It is straightforward to calculate the
extrinsic curvature of this slice in terms of the perturbations of the
spacetime metric. In the RW notation (but not the RW gauge), the
results are
\begin{eqnarray}
{\cal A}&=&\Phi^{-2}\left[
(1-2M/r)^{-1/2}H_1'-\textstyle{\frac{1}{2}}(1-2M/r)^{-3/2}\dot{H}_2+
(1-2M/r)^{-3/2}(M/r^2)H_1
\right]\label{KdecompA}\\
{\cal B}&=&\Phi^{-2}\textstyle{\frac{1}{2}}(1-2M/r)^{-1/2}
\left[
H_1 +h_0^\prime-\dot{h}_1-2h_0/r
\right]\\         
{\cal K}&=&\Phi^{-2}
\left[r(1-2M/r)^{1/2}H_1
-\textstyle{\frac{1}{2}}r^2(1-2M/r)^{-1/2}\dot{K}
\right]\\
{\cal G}&=&
\Phi^{-2}(1-2M/r)^{-1/2}
\left[h_0
-\textstyle{\frac{1}{2}}r^2\dot{G}
\right]\label{KdecompG}\ ,
\end{eqnarray}
where prime indicates derivate with respect to $r$, and dot with
respect to $t$.

To find the initial value of $\dot{\psi}$, we use the prescription
given by Abrahams and Price\cite{ap1}, and we 
write
\begin{equation}\label{formpsi}
\psi={\cal Q}\left\{
\delta g_{ij}, \delta g'_{ij}
\right\}\ ,
\end{equation}
in which ``${\cal Q}$'' represents Moncrief combination\cite{moncrief} of
perturbations of the metric $\delta g_{ij}$, and of the $r$ derivative
of $\delta g_{ij}$. The initial value of $\dot{\psi}$ is
then the initial value of
\begin{equation}\label{formpsidot}
\dot{\psi}=-2{\cal Q}\left\{(1-2M/r)^{1/2}
\delta K_{ij}, \left((1-2M/r)^{1/2}\delta K_{ij}
\right)'\right\}\ .
\end{equation}

By comparing Eqs.\ (\ref{rwform}) and 
(\ref{Knotation})-(\ref{Knot_last})
we find that the substitutions needed are:
\begin{eqnarray}\label{substits}
(1-2M/r)^{-1}H_2&\longrightarrow&\Phi^{-2}{\cal A}
\left(\partial \bar{r}/\partial r
\right)^2\sqrt{1-2M/r}\\
K&\longrightarrow&\Phi^{-2}{\cal K}\left(\bar{r}/r
\right)^2\sqrt{1-2M/r}\\
G&\longrightarrow&\Phi^{-2}{\cal G}\left(\bar{r}/r
\right)^2\sqrt{1-2M/r}\\
h_1&\longrightarrow&\Phi^{-2}{\cal B}
\left(\bar{r}/r
\right)\sqrt{1-2M/r}\label{substits_end} .
\end{eqnarray}
With these substitutions made in Eq.\ (\ref{psidef}) we get a specific
expression for $\dot{\psi}$, as a function of $r$, on the
hypersurface.  An independent derivation of the result starts with the
expression
\begin{equation}\label{psidoteq}
\dot{\psi}(r,t)=\frac{r}{{\lambda}+1}\left[
    \dot{K}+\frac{r-2M}{{\lambda}r+3M}\left\{ \dot{H}_2-r\partial \dot{K}/
\partial r
    \right\} \right]
+(r-2M)\left(r^2\partial \dot{G}/\partial r-2\dot{h}_1
\right)\ ,
\end{equation}
where dots indicate $\partial/\partial t$. 
With 
Eqs.\ (\ref{KdecompA})--(\ref{KdecompG}).
the right hand side of Eq.\
(\ref{psidoteq}) can be reexpressed in terms of ${\cal A,B,K,G}$. We
have checked that the result is identical to that found from
Eqs.\ (\ref{formpsidot})--(\ref{substits_end}), i.e.
\begin{eqnarray}\label{psipunto}
\dot{\psi}(r,t)=-\frac{2\bar{r}\sqrt{1-2M/r}}{(\lambda+1)r^2}\Biggr\{
\bar{r}^2{\cal G}+\frac{\sqrt{1-2M/r}}
{(2\lambda+6M/r)}\bigg[&-&4(\lambda+1)(\bar{r}^2{\cal B})
+{M\over r}\bigg(\frac{1}{\sqrt{1-2M/r}}+\sqrt{r\over\bar{r}}\bigg)
(\bar{r}^2{\cal A})\nonumber\\
&+&\bar{r}\partial_{\bar{r}}(\bar{r}^2{\cal A})
\bigg]\Biggr\}\ .
\end{eqnarray}

In describing the general choice of $\dot{\psi}$ on a constant $t$
hypersurface, it is useful to review the equivalent question for
$\psi$.  If we restrict our choice to a conformally flat 3-geometry,
then the hamiltonian constraint, and the constraint of conformal
flatness, reduces the specification of the initial geometry, for each
multipole, to a single ordinary differential equation, Eq.\ 
(\ref{constraint}) with $H_2=K$.  The only choice to be made is the
constant, specifying how much of the physically well-behaved
homogeneous solution (\ref{homog}) is to be added to a particular
solution; specifying this constant is equivalent to choosing between,
for example, the nonsymmetric (Brill-Lindquist) solution of Eq.\ 
(\ref{BLPhi}) and the reflection symmetric (Misner) solution of Eq.\ 
(\ref{MisPhi}).  If, on the other hand, we do not restrict ourselves
to a conformally flat 3-geometry, we have a functional degree of
freedom in the initial geometry. (Equivalently, in the RW gauge we
have the freedom to specify $H_2-K$ for each multipole.) Whether or
not we restrict to a conformally flat 3-geometry, the Moncrief
wavefunction $\psi$, defined in Eq.\ (\ref{psidef}), is fixed once the
3-geometry is specified.  Furthermore, $\psi$, is totally gauge
invariant.

For $\dot{\psi}$ the situation is closely parallel. Analogous to the
condition on 3-geometry, that it be conformally flat, we now have the
special condition that the extrinsic curvature be purely
``longitudinal,'' i.e., derivable from a potential according to Eqs.\ 
(\ref{KvsKhat}) and (\ref{KfromW}).  For our axisymmetric infall, the
mathematical problem of finding the $\widehat{W}_i$ potential, for
each multipole, is the problem of finding the two components 
$\widehat{W}_{{\bar{r}}}$ and $\widehat{W}_\theta$. These can be expanded in
multipoles and written as:
\begin{equation}\label{Wdecomp}
\widehat{W}_{\bar{r}}=\sum_\ell A_\ell(\bar{r})Y_{\ell0}\hspace*{.9in} 
\widehat{W}_\theta=\sum_\ell B_\ell(\bar{r})
\partial Y_{\ell0}/\partial\theta\ .
\end{equation}
The nontrivial momentum constraints, equivalent to the $t\theta$ and
$tr$ components of the einstein equations, then constitute two second
order differential equations for $A_\ell(\bar{r})$ and $B_\ell(\bar{r})$.
A particular solution to the momentum constraints is, of course, given
by the fundamental solution in Eq.~(\ref{Kfund}). This solution
corresponds to the $\widehat{W}^i$ multipoles
\begin{equation}\label{Afund}
A^{\rm basic}_\ell(\bar{r})=\sqrt{\frac{4\pi}{2\ell+1}}\,\frac{P}{4\bar{r}}
\,\left[\frac{\ell(\ell+7)}{2\ell-1}q^{\ell-1}
+\frac{(\ell+1)(6-\ell)}{2\ell+3}q^{\ell+1}
\right]
\end{equation}
\begin{equation}\label{Bfund}
B^{\rm basic}_\ell(\bar{r})=\sqrt{\frac{4\pi}{2\ell+1}}\,\frac{P}{4}
\,\left[\frac{(8-\ell)}{2\ell-1}q^{\ell-1}
+\frac{(\ell-6)}{2\ell+3}q^{\ell+1}
\right]\ .
\end{equation}
The above expressions are valid only for $q\equiv
\bar{z}_p/\bar{r}\leq1$.  For $q>1$ the replacement $\ell\rightarrow
-\ell-1$ must be made inside the square brackets.

One finds that there are only two solutions well behaved at infinity,
and that all solutions are well behaved at the horizon.  There are
then two well-behaved homogeneous solutions of the momentum
constraints that can be added to the particular solution of Eqs.\ 
(\ref{Afund}) and (\ref{Bfund}). 

Homogeneous longitudinal solution to the momentum constraints can be
found directly from the differential equations. Alternatively one can
construct those from known solutions. One such solution is an ``image
solution,'' the basic solution given in Eq.\ (\ref{Wfund}), but with
$\bar{z}_p$ replaced by the location $\bar{z}_{\rm image}<M/2$, so
that the position of the $\bar{z}=\bar{z}_{\rm image}$ singularity is
inside the horizon of the background hole.  (If the choice $z_{\rm
  image}=(M/2)/\bar{z}_p$ is made, then the singularity is the
``image'' of the particle under inversion through the spherical
surface at $\bar{r}=M/2$.) The image solution is, then, the following
simple modification of Eq.\ (\ref{Afund}) and (\ref{Bfund}):
\begin{equation}\label{Aimage}
A^{\rm image}_\ell(\bar{r})=\sqrt{\frac{4\pi}{2\ell+1}}\,\frac{P}{4\bar{r}}
\,\left[\frac{\ell(\ell+7)}{2\ell-1}q_{\rm image}^{\ell-1}
+\frac{(\ell+1)(6-\ell)}{2\ell+3}q_{\rm image}^{\ell+1}
\right]
\end{equation}
\begin{equation}\label{Bimage}
B^{\rm image}_\ell(\bar{r})=\sqrt{\frac{4\pi}{2\ell+1}}\,\frac{P}{4}
\,\left[\frac{(8-\ell)}{2\ell-1}q_{\rm image}^{\ell-1}
+\frac{(\ell-6)}{2\ell+3}q_{\rm image}^{\ell+1}
\right]\ ,
\end{equation}
Here $q_{\rm image}\equiv \bar{z}_{\rm image}/\bar{r}$ and is always
less than unity for $\bar{r}$ outside the horizon (i.e., $\bar{r}>M/2$).
Similarly, we can find the multipole decomposition of the
corresponding extrinsic curvature, simply by replacing $q$ by
$q_{\rm image}$ in Eqs.\ (\ref{fundCals}) --(\ref{lastbasic}). 

A second longitudinal solution can be constructed from the ``$\alpha$''
solution given by Bowen and York\cite{bowenyork,bowen} and elsewhere:
\begin{equation}\label{Kalpha}
\widehat{K}^{\alpha}_{ij}=\frac{3}{2\bar{\rho}^4}\left[
P_i n_j+P_j n_i+\left(\gamma^{\rm flat}_{ij}-5n_i n_j
\right)P^k n_k
\right]\ 
\end{equation}
The equivalent  $\widehat{W}_i$ vector is
\begin{equation}\label{Walpha}
\widehat{W}^{\alpha}_i=
-\frac{1}{4\bar{\rho}^3}\left[P_i-3n_iP^k n_k\right]\ .
\end{equation}
This solution has been of interest in connection with inversion
symmetry, but we use it here with no prejudice about symmetry of the
solution under inversion. To make the solution of Eqs.\ (\ref{Kalpha})
and (\ref{Walpha}) a homogeneous solution we need only place the
$\bar{\rho}=0$ singularity inside the horizon, say on the $\bar{z}$ axis at
$\bar{z}_{\alpha}<M/2$. 

The multipole decomposition of the $\alpha$ solution, in the notation of
 Eqs.\ (\ref{Knotation}) --
(\ref{Knot_last}) is
\begin{eqnarray}\label{alphaCals}
{\cal A}^{\alpha}_\ell&=&\sqrt{\frac{ 4\pi }{2\ell+1}}\, \frac{P}{2\bar{r}^4}\, 
\ell(\ell+1)(\ell+2)q_{\alpha}^{\ell-1}
\\
{\cal B}^{\alpha}_\ell&=&-\sqrt{\frac{ 4\pi }{2\ell+1}}\,
\frac{P}{2\bar{r}^3}\,\ell(\ell+2)q_{\alpha}^{\ell-1}
\\
{\cal G}^{\alpha}_\ell&=&\sqrt{\frac{ 4\pi }{2\ell+1}}\, \frac{P}{2\bar{r}^4}\,
\ell q_{\alpha}^{\ell-1}
\\
{\cal K}^{\alpha}_\ell&=&-\sqrt{\frac{ 4\pi }{2\ell+1}}\, \frac{P}{2\bar{r}^4}\,
\ell(\ell+1) q_{\alpha}^{\ell-1}\ ,
\label{lastalpha}
\end{eqnarray}
where $q_\alpha\equiv \bar{z}_\alpha/\bar{r}$ and is always less
than unity for $\bar{r}$ 
outside the horizon. For the $\alpha$ solution, the multipole decomposition
of the $\widehat{W}_i$ vector, in the notation of Eq.\ (\ref{Wdecomp}), is
\begin{equation}\label{Aalph}
A^{\alpha}_\ell(\bar{r})=-\sqrt{\frac{4\pi}{2\ell+1}}\,\frac{P}{4\bar{r}^3}
\ \ell(\ell+1)q_\alpha^{\ell-1}
\end{equation}
\begin{equation}\label{Balph}
B^{\alpha}_\ell(\bar{r})=\sqrt{\frac{4\pi}{2\ell+1}}\,\frac{Pa}{4\bar{r}^2}
\ \ell q_\alpha^{\ell-1}\ .
\end{equation}

The general longitudinal solution for the extrinsic curvature, for
each multipole, is a superposition of three contributions: (i) the
particular solution given by Eqs.\ (\ref{fundCals})-(\ref{lastbasic}),
(ii) the homogeneous ``image'' solution corresponding to Eqs.\ 
(\ref{Aimage}), (\ref{Bimage}), for some $\ell$-dependent choice of
$\bar{z}_{\rm image}$, multiplied by an arbitrary, $\ell$-dependent
constant, and (iii) the homogeneous ``$\alpha$'' solution in Eqs.\ 
(\ref{Kalpha}) -- (\ref{Balph}), for some $\ell$-dependent choice of
$\bar{z}_{\alpha}$, multiplied by an arbitrary, $\ell$-dependent constant.
One is led to ask what the ``proper'' choice is for the general
longitudinal solution. A somewhat broader question is whether the
initial data should be chosen to be longitudinal.

In the initial data solutions generated for use in numerical
relativity, the choice has often been made to take the solution to be
inversion symmetric\cite{cook}. (That solution turns out not, in fact,
to be longitudinal.) The criterion used in numerical relativity has
been definiteness and the simple-to-implement inner boundary
condition provided by inversion symmetry. Here our criterion shall be
quite different: we study what choice of hypersurface data is the best
representation of the solution that evolves from an earlier
astrophysical configuration. More specifically, we can start a
particle falling from some large radius $r_0$, and numerically evolve
$\psi$ to a hypersurface at some later time. We can then study the
agreement between that evolved data and the data given by some
prescription for assigning hypersurface data to represent the falling
particle.

It should be understood that in making this comparison (i.e., evolved
data vs. prescribed data) there are no issues of ``choice'' involving
the slicing of spacetime. The Moncrief wave variable $\psi$ is
invariant with respect to perturbative changes in slicing (as well as
shifts and diffeomorphism), so that our comparison of evolved $\psi$
and prescribed $\psi$ involves no gauge related ambiguities.

\subsection{Choice of hypersurface data}

In our discussion above of initial data we found the following: (i) If we
restrict ourselves to conformally flat data, we must choose a single
constant, for each multipole, to specify our solution for $\psi$; if
we do not restrict to conformally flat data there is an unconstrained
function of $r$ for each initial multipole of $\psi$. (ii) If we
restrict our choice to longitudinal initial extrinsic curvature, then
we must specify two constants for each multipole, in order to fix the
initial $\dot{\psi}$; if we do not restrict to initial longitudinal 
extrinsic curvature, then there is an unconstrained function of $r$
for each multipole of $\dot{\psi}$ (in our case of axially symmetric
collision).

Our goal below will be to see whether there is a simple physical
choice that can be made for the prescribed hypersurface data that
agrees well with the data evolved from $t=0$. We shall try to reach
that goal by restricting our prescription to conformally-flat,
longitudinal (hereafter CFL) data.  The question we shall be asking,
therefore, is whether there exists a choice of constants specifying
the CFL solution for the particle on  a hypersurface so that it agrees well
with an evolved solution. Such a choice of constants could be
determined by looking for some kind of ``best fit'' of the CFL
solution to the evolved solution, but of much greater interest is a
physically-based prescription for the choice of constants which can be
used as a best guess of appropriate initial data when an evolved
solution is not available.

The rule we shall use for the specification of the CFL $\psi$ is based
on a simple physical consideration. We fix the constant
in the CFL solution for $\psi$ by demanding that the value of $\psi$
at the horizon be ``frozen,'' i.e., fixed at the value it has on the
initial hypersurface $t=0$. This is in accord with the idea of
dynamics freezing at the horizon, but it is better to view it as a
consequence of the wave equation Eq.\ (\ref{rtzerilli}), and its
horizon limit at $r^*\rightarrow-\infty$.  The initial data at $t=0$
is a smooth function of $\bar{r}$ near the horizon,
and hence $\psi$ becomes constant as
$r^*\rightarrow-\infty$. But the potential decreases as $e^{r^*/2M}$
and can be ignored near the horizon, so $\psi=$\, constant is a static
solution of the wave equation. The value of $\psi$ at the horizon will
therefore be unchanging in time. We shall see, in the numerical
results presented below, that the evolved data demonstrate this
freezing.

It is straightforward to use this condition of freezing at the horizon
to fix the constant multiplier of the homogeneous solution in Eq.\ 
(\ref{homog}), and the consequence is interesting. If we start with a
BL solution (i.e., the solution in Eq.\ (\ref{particular}) with the
particle position at $\bar{z}_p=\bar{r}_0$) then the ``horizon
frozen'' solution, on a later hypersurface, turns out to be a
superposition of the following simple solutions: (i) the BL solution
corresponding to the the particle at its correct position $\bar{z}_p$
at time $t$, (ii) a solution for a particle inside the horizon located
at $\bar{z}_{\rm image}=(m/2)^2/\bar{z}_p$, so that when added to the
BL solution the total value of $\psi$ vanishes at the horizon, and
(iii) a solution corresponding to a particle at the image location of
the original position $\bar{z}=(M/2)^2/\bar{r}_0$. The magnitude of the
mass parameter for this last image gives a $\psi$ at the horizon equal
to the frozen value.

The way in which we fix the constants for $\dot{\psi}$ is a more
complicated, less elegant, implementation of the same basic physical
idea of looking at the nature of solutions of Eq.\ (\ref{rtzerilli}),
near the horizons.
At $t=0$ our initial data is taken to be stationary, and of the form 
\begin{equation}
\psi=a +b\bar{r}+c\bar{r}^2+\cdots\ .
\end{equation}
Since $\bar{r}$ is related to $r^*$ by
\begin{equation}\label{rstvsrbar}
\bar{r}=M/2+Me^{-1/2}e^{r^*/4M}+Me^{-1}e^{r^*/4M}+{\cal O}(e^{3r^*/4M})\ ,
\end{equation}
it follows that our initial solution has the horizon limit
\begin{equation}\label{inithordata}
\psi =\tilde{a}+\tilde{b}e^{r*/4M}
+\tilde{c}e^{r*/2M}\ ,\hspace*{20pt}\dot{\psi}=0\ .
\end{equation}
In the horizon limit, the  potential $V_\ell$, in Eq.\ (\ref{rtzerilli}),
takes the form
\begin{equation}
V_\ell=e^{r^*/2M}V_o +{\cal O}(e^{r^*/M})\ ,\hspace*{20pt}
V_o=\frac{4\lambda^2+4\lambda+3}{(2\lambda+3)4M^2e}
\end{equation}
where $V_o$ is a constant.
In the horizon limit we are, then, solving the problem
\begin{equation}\label{rtzhorizon}
-\frac{\partial^2\psi}{\partial t^2}
+\frac{\partial^2\psi}{\partial r*^2}-
e^{r^*/2M}V_o\psi=0\ ,
\end{equation}
with the initial data given by Eq.\ (\ref{inithordata}).
It is simple to show that the solution to this problem
is
\begin{equation}\label{nrhorizon}
\psi=\tilde{a}+\tilde{b}\cosh{(t/4M)}e^{r^*/4M}
+\left[\left(\tilde{c}-V_o(2M)^2\tilde{a}\right)\cosh{(t/2M)}
+V_o(2M)^2\tilde{a}\right]e^{r^*/2M}\ .
\end{equation}
We can now take the time derivative of this near-horizon solution, evaluate
it on the hypersurface at time $t$, and find
\begin{equation}\label{nrhorizondot}
\dot{\psi}|_{t}=(\tilde{b}/4M)\sinh{(t/4M)}e^{r^*/4M}
+\left(\tilde{c}-V_o(2M)^2\tilde{a}\right)\sinh{(t/2M)}e^{r^*/2M}\ .
\end{equation}
This result can be compared with the Taylor expansion around
$\bar{r}=M/2$ of the ``basic,'' ``image,'' and ``$\alpha$''
longitudinal solutions, and the two constants fixing the solution
thereby determined.

{}From Eq.\ (\ref{nrhorizon}) it can be immediately seen that $\tilde{a}$
does not depend on time. This is equivalent to the condition of ``freezing''
on the horizon. The amplitude $\Gamma$ by which we have to multiply the
homogeneous solution\ (\ref{homog}) is
\begin{equation}\label{uno}
\Gamma_{homog}(\bar{z}_p)={m_2(r_0)\over m_2(z_p)}\left({\bar{z}_p\over
\bar{r}_0}\right)^{l+1}\left(1+\Gamma_{homog}(r_0)\right)-1~,
\end{equation}
where $\Gamma_{homog}(r_0)$ represents the arbitrariness in choosing
the data on the initial hypersurface $t=0$. For instance,
$\Gamma_{homog}(r_0)=0,1,-1$ represents respectively BL, Misner, and
antisymmetric data.

The amplitude of the ``image'' term in the extrinsic curvature is
the following
\begin{eqnarray}\label{dos}
\Gamma_{image}=&&\Delta^{-1}\Biggr\{{4q_3^{l+2}q_4^{l-1}\over M^2}
\left[{6l(4l+1)\over2l+3}\right]\nonumber\\
&&+{16m_p(r_0)(2l+1)q_4^{2l}\over M^2P}\biggr[4l
(1-\Gamma_{homog}(r_0))\sin(t/4M)\nonumber\\
&&+{(1+\Gamma_{homog}(r_0))(5-l-3l^2-4l^3-2l^4)
\over(l+1)(l^2+l+1)}\sin(t/2M)\biggr]\Biggr\}~,
\end{eqnarray}
where
\begin{equation}\label{tres}
\Delta={4q_3^{l-1}q_4^{l-1}\over M^2}\left({2l\over2l+3}\right)
\left[-(5l+1)(2l+3)+(l-6)(2l+1)q_3^2\right]~,
\end{equation}
$\bar{z}_{image}=(M/2)^2/\bar{z}_p$, $q_3=M/(2\bar{z}_p)$, and
$q_4=M/(2\bar{r}_0)$.

Finally, the amplitude of the $\alpha$ term in the extrinsic curvature
(see Eq.\ (\ref{Kalpha})) is
\begin{eqnarray}\label{cuatro}
\Gamma_{\alpha}=&&\Delta^{-1}\Biggr\{2q_3^{2l-1}\biggr[
{-(l+7)(5l+1)\over2l-1}+{(2l+1)(26l^2+26l-27)\over(2l-1)(2l+3)}q_3^2-
{(l-6)(5l+4)\over2l+3}q_3^4\biggr]\nonumber\\
&&+{16m_2(r_0)(1-\Gamma_{homog}(r_0))(2l+1)q_3^{l-1}q_4^{l+1}\over P}
\biggr[-{5(l-2)\over2l-1}+{(l-6)\over2l+3}q_3^2\biggr]\sin(t/4M)\nonumber\\
&&+{4m_2(r_0)(1+\Gamma_{homog}(r_0))q_3^{l-1}q_4^{l+1}\over
l(l+1)(l^2+l+1)P}\biggr[-{10l^2+9l+8\over2l-1}+{(l-6)(2l+1)\over2l+3}
q_3^2\biggr]\sin(t/2M)\Biggr\}~,
\end{eqnarray}
where $\bar{z}_{\alpha}=(M/2)^2/\bar{r}_0$.

Note that this choice of the amplitudes we assign to the homogeneous
solutions fixes all three constants in the general solution, but this is not
the only way we can fix them. We could have chosen, for instance,
matching the behavior of the evolved $\psi$ for large $r$. This, however,
leads to a less successful approximation.

\section{Results}
\subsection{Numerical method}

In this section we describe the algorithm used to integrate the wave
equation (\ref{rtzerilli}) numerically. While the left hand side of
the equation is straightforward to integrate, the source given by
Eq.\ (\ref{rtsource}) contains terms with a Dirac's delta and its
derivative. Since we have not found in the literature a discussion of
the numerical treatment of such sources, we shall describe the method
in some detail.

We have found it convenient to use a numerical scheme with step sizes
$\Delta t=\frac{1}{2}\Delta r^*\equiv h$, and with a staggered
grid. As Fig.\ \ref{fig:rombo}  shows, this method connects
points along lines of constant ``retarded time'' $u\equiv t-r^*$ and
``advanced time'' $v\equiv t+r^*$.  On this grid we have implemented a
finite difference algorithm for evolving $\psi$ with errors of order
$h^2$; that is, at a given value of $t$ and $r*$ a measure of the
solution error varies as $h^2$.  This method has proven to be easy
to implement and quite accurate.

To derive our difference scheme we start by integrating 
Eq.\ (\ref{rtzerilli}) over the cell of our numerical grid shown in 
Fig.\ \ref{fig:rombo}. We use the notation 
\begin{equation}
\int\int dA=
{\int\int}_{\rm cell}dt\,dr^*=\int^{u+h}_{u-h}du~\int^{v+h}_{v-h}dv\ .
\end{equation}
Applied to the derivative terms in Eq.\ (\ref{rtzerilli}) this gives:
\begin{eqnarray}\label{intderivs}
&&
\int\int dA \left\{-\partial_t\partial_t\psi+\partial_r^*\partial_r^*\psi
\right\}=\int\int dA\left\{-4\partial_u\partial_v\psi
\right\}=\nonumber\\
&&
-4\left[\psi(t+h,r^*)+\psi(t-h,r^*)-\psi(t,r^*+h)-\psi(t,r^*-h)\right]\ .
\end{eqnarray}
Note that this result is exact; it contains no truncation errors.

We next consider the integration of the potential term over the 
cell. If the cell is one with no source term contribution, then 
we can use
\begin{eqnarray}\label{cellwosource}
&&\int\int dA\left\{-V\psi\right\}=
-h^2\left[V(r^*)\psi(t+h,r^*)+V(r^*)\psi(t-h,r^*)+\right.\nonumber\\
&&\left. V(r^*+h)\psi(t,r^*+h)+
 V(r^*-h)\psi(t,r^*-h)\right]+{\cal O}(h^4)\ .
\end{eqnarray}
The $h^4$ order error in a generic cell is equivalent to an overall
${\cal O}(h^2)$ error in $\psi$.

The result in Eq.\ (\ref{cellwosource}) assumes that $\psi$ is smooth
in the grid cell. It cannot be applied to those cells through which
the particle worldline passes, since $\psi$ is discontinuous across
the worldline. For such cells we first obtain the coordinates
$(r^*_1,t_1)$ of the point where the particle enters the cell and
$(r^*_2,t_2)$ where the particle leaves it (see Fig.\ 
\ref{fig:rombo}).  Next, we divide the total area of the cell,
$(4h^2)$, into four subareas defined as follows: $A_2$ is the part of
the area of the rhomb below $t=t_1$, $A_3$ is the part of the area of
the rhomb over $t=t_2$, $A_1$ is the remaining area to the left of the
particle's trajectory, and $A_4$ is the remaining area to the right.

The integral of the $V\psi$ term over the area of the cell is 
approximated by the sum of this function evaluated on the corners
of the cell multiplied by the corresponding subarea $A_i$. 
This gives us
\begin{equation}\label{cellwsource}
\int\int dA\left\{-V\psi\right\}= -V(r*)
\left[
\psi(t+h,r^*)A_3
+\psi(t,r^*+h)A_4
+\psi(t,r^*-h)A_1
+\psi(t-h,r^*)A_2
\right]\ .
\end{equation}
The truncation error in each such cell is of order $(h^3)$,
just enough to have quadratic convergence, since only one cell with
the particle has to be evaluated per time step.

For cells through which the worldline passes, the integral of the 
source term in Eq.\ (\ref{rtsource}) must be evaluated.
As a convenience in discussing the numerical approximation of this term
Eq.\ (\ref{rtsource}), we introduce the notation
\begin{equation}\label{fuente}
{\cal S}=G(t,r)\delta[r-r_p(t)]+F(t,r)\delta'[r-r_p(t)]\ .
\end{equation}
The integration over the cell, when done with due regard to the boundary
terms generated by the 
$\delta'[r-r_p(t)]$, gives
\begin{eqnarray}\label{integral}
\int\int{\cal S} dA=&&2\int_{t_1}^{t_2}d
t\left[{G\left(t,r_p(t)\right)\over1-2M/r_p(t)}-
\partial_r\left({F(t,r)\over1-2M/r}\right)
\biggr\vert_{r=r_p(t)}\right]
\nonumber\\
&&
\pm2{F\left(t_1,r_p(t_1)\right)\over\left(1-2M/r_p(t_1)\right)^2}
\left(1\mp\dot r_p^*(t_1)\right)^{-1}\nonumber\\
&&
\pm2{F\left(t_2,r_p(t_2)\right)\over\left(1-2M/r_p(t_2)\right)^2}
\left(1\pm\dot r_p^*(t_2)\right)^{-1}.
\end{eqnarray}
The $\int\,dt$ integral in the first term can be performed to any
precision since $F$ and $G$ are known functions. For our goal of
quadratic convergence, a trapezoidal approximation for the integration
is adequate.  In the second term the upper (lower)
sign is for particles entering the cell from the right (left), or
equivalently for $r^*_1> r^*$ ($r^*_1< r^*$). In the same way, in the
third term the upper (lower) sign is for particles leaving the cell to
the right (left), or equivalently $r^*_2> r^*$  $(r^*_2< r^*)$.

When the form of $F$ and $G$, given in Eq.\ (\ref{rtsource}), are used
in Eq.\ (\ref{integral}) the result is
\begin{eqnarray}\label{FyG}
&&{F(t,r_p(t))\over\left(1-2M/r_p(t)\right)^2}={\kappa\epsilon_0^{-1}(r_p-2M)
\over(\lambda+1)(\lambda r_p+3M)}~;~~\epsilon_0\equiv\sqrt{1-2M/r_0},
\nonumber\\
&&
{G(t,r_p(t))\over1-2M/r_p(t)}-\partial_r\left({F(t,r)\over1-2M/r}\right)
\biggr\vert_{r=r_p(t)}
={\kappa\epsilon_0^{-1}(1-2M/r_p)\over(\lambda+1)r_p(\lambda r_p+3M)^2}
\left[3M^2-4\lambda Mr_p-\lambda(\lambda+1)r_p^2-6M(1-\epsilon_0^2)r_p\right],
\nonumber\\
&&
\dot r_p^*(t)=-\epsilon_0^{-1}\sqrt{2M/r_p-2M/r_0}\ .
\end{eqnarray}

Our numerical scheme, for cells through which the particle worldline
does not pass, is to solve for $\psi(t,r^*)$, using Eq.\ 
(\ref{intderivs}) and Eq.\ (\ref{cellwosource}) in the integral of
Eq.\ (\ref{rtzerilli}). For cells containing the worldline, Eq.\ 
(\ref{intderivs}), Eq.\ (\ref{cellwsource}), Eq.\ (\ref{integral}) and
Eq.\ (\ref{FyG}) are used. In summary, the evolution algorithm we use is
\begin{equation}
\psi(t+h,r^*)=-\psi(t-h,r^*)+\left[\psi(t,r^*+h)+\psi(t,r^*-h)\right]\left[1-
\frac{h^2}{2}V(r^*)\right]~,
\end{equation}
for cells not crossed by the particle, and
\begin{eqnarray}
\psi(t+h,r^*)=&&-\psi(t-h,r^*)\left[1+\frac{V(r^*)}{4}(A_2-A_3)\right]
+\psi(t,r^*+h)\left[1-\frac{V(r^*)}{4}(A_4+A_3)\right]\nonumber\\
&&
+\psi(t,r^*-h)\left[1-\frac{V(r^*)}{4}(A_1+A_3)\right]-
\frac{1}{4}\int\int{\cal S} dA~,
\end{eqnarray}
for the cells that the particle does cross.

The above equations cannot, however, be used to
initiate the evolution off the first hypersurface.  If $t_0$ denotes the
time at which we have the initial data, we lack the values
$\psi(t_0-h)$, necessary to apply the evolution algorithm.  We can,
however, use a Taylor expansion to write
\begin{equation}\label{punto}
\psi(t_0-h,r^*)=\psi(t_0+h,r^*)-2h\partial_t\psi(t_0,r^*)+{\cal O}(h^3)\ .
\end{equation}
The right hand side can be used in place of $\psi(t_0-h,r^*)$ in the
application of the algorithm to evolve off the first hypersurface.
It is important to note that this substitution is valid only if
$\psi(t,r^*)$ is not singular between $t=t_0-h$ and $t=t_0-h$. This
requires that the particle worldline not cross the vertical line at
$r^*$ between $t=t_0-h$ and $t=t_0+h$. In setting up the computational
grid, we have been careful always to avoid such a crossing.

The numerical method used here, evolving initial data for a partial
differential equation on a staggered grid, has little in common with
the transform method used in Paper I, in which we studied only
momentarily stationary initial conditions.  A comparison of the two
methods (in the case of momentarily stationary initial data) provided
a powerful check of both methods as well as insights into the relative
efficiency and accuracy of the methods.  The agreement of the two
approaches turned out to be excellent. For the goal of producing an
evolved waveform, the numerical evolution method was found to be
faster to a transform method by orders of magnitude, and to give more
accurate results.

\subsection{Numerical results}

In this section we present computed results for the infall of a
particle, starting from rest at initial coordinate position $r_0$.
The particle trajectory is a radial geodesic as described by Eq.\ 
(\ref{time}).  By ``$t=0$,'' we shall always mean the hypersurface at
which the particle was at rest. We shall consider two values of
$r_0/2M$: both 15 and 1.5.  The former is a starting point where the
influence of the background hole is reasonably weak. (The
gravitational redshift, for example, is around only 3\%.) This
represents, then, an astrophysical starting point for which a
Newtonian description would be a good approximation. The second
choice, $r_0/2M=1.5$ is not a reasonable astrophysical starting point,
but complements the first choice of $r_0$, magnifies certain effects,
and is useful for exploration and illustration.

We will be interested in Cauchy data, both $\psi$ and $\dot{\psi}$, on
subsequent $t=$\, constant hypersurfaces. (We repeat here an important
feature of the formalism: the gauge invariant information $\psi$ and
$\dot{\psi}$ is unaffected by a change in slicing. When we say,
therefore, a $t=$\, constant hypersurface, we mean only $t=$\,
constant to zero order in the particle mass.) There are several ways
in which we can specify which later slice we are considering. (i) We
could specify the Schwarzschild coordinate time $t$ for this
hypersurface. (ii) We could specify the location of the infalling
particle corresponding to time $t$; for this purpose we use the
notation $r_p$ (or $r_p^*$, the $r^*$ equivalent). Note that $r_p$ is
related to the position parameter $\bar{z}_p$ by
$r_p=\bar{z}_p(1+M/2\bar{z}_p)^2$.  (iii) We could specify the parameter
$P$ describing the particle's momentum at time $t$. (This is the $P$
parameter that enters into the Bowen-York description of extrinsic
curvature; see the discussion in Sec.\ II relating this parameter to
the particle's 4-momentum components. A picture of a sliced
Schwarzschild spacetime, illustrating the particle trajectory and the
labeling of the spacetime, is shown in Fig.\ \ref{fig:traj}.

In our results the crucial concept is the difference between
``evolved'' data and ``prescribed'' data.  By ``evolved'' values of
$\psi$ and $\dot{\psi}$ we shall mean the values that are found on
some hypersurface after numerical evolution forward in time from the
$t=0$ original hypersurface. By ``prescribed data'' we shall mean data
that is chosen according to one of the prescriptions of Sec.\ II.
Prescribed data, for our purposes here, is always chosen from the set
of possibilities that we describe as conformally flat, longitudinal.
This means that the 3-geometry is conformally flat, and the extrinsic
curvature is ``longitudinal,'' in the sense defined by
York\cite{yorkjmp}. For $\psi$, specific choices that we have
described in Sec.\ II include: (i) BL data, the simplest choice for
$\psi$, (ii) Misner data, data that is symmetric with respect to
reflection under $\bar{r}\rightarrow(M/2)^2/\bar{r}$ in the
conformally related flat space, (iii) ``frozen'' data, the conformally
flat solution of the hamiltonian constraint that has the value of
$\psi$ at the horizon fixed to be the horizon value on the original
($t=0$) hypersurface.  Prescribed choices for $\dot{\psi}$ include (i)
the basic Bowen-York \cite{bowenyork} solution, the simplest
longitudinal solution, and (ii) ``horizon matched'' data, the
longitudinal solution with the form of $\dot{\psi}$ (more
specifically, the first and second derivatives with respect to $r^*$)
matched to the evolved solution at the horizon limit.

We first present comparisons of radiated energy. In Fig.\ 
\ref{fig:energyAa} we show the quadrupole energy $E_2$ radiated during
the infall of the particle from $r_0=1.5(2M)$.  On the original $t=0$
hypersurface, the form of $\psi$ is taken to be the Misner solution.
(Since the particle is momentarily stationary at $t=0$, we have
$\dot{\psi}=0$ on this hypersurface, of course.)  The plot shows the
energy radiated from the particle for times after the particle is at a
location $r^*=r_p^*$. These energies are computed from the solution
that evolves from the data on the constant time hypersurface labeled
$r_p^*$.  The ``true'' data are the data that evolved from the
$t=0$ original hypersurface on which the particle was momentarily
stationary.  The energy computed in this way is the ``true'' total
quadrupole energy emitted during infall from $15(2M)$, and has the
value $8.1\times10^{-3}m_0^2/(2M)$.

The plot shows the energy generated when the true (i.e., evolved)
$\psi$ and $\dot{\psi}$, on a hypersurface, are replaced with
prescribed data appropriate on that hypersurface for the position and
momentum of the geodesically falling particle.  To find the points for
the dotted curve, for example, on each hypersurface, position and
momentum of the infalling particle were calculated and used to
generate Misner data for $\psi$, and Bowen-York data for $\dot{\psi}$,
on that hypersurface. These prescribed data were then numerically
evolved forward in time, and the resulting radiation to a distant
observer was computed. This radiated energy is presented as a function
of the hypersurface label $r_p^*$.  Similarly, the solid curve shows
the result of replacing the true hypersurface data with frozen data
for $\psi$ and horizon-matched data for $\dot{\psi}$. The energies of
the dashed curve are the result of replacing the $\dot{\psi}$ data
with horizon-matched data, and retaining the true $\psi$ data.

These results show that the ``frozen-matched'' prescription are in
slightly better agreement with the correct radiated energy than are
the other choices. The choice of frozen-matched conditions is also
more justifiable, since it is based on a physical consideration.
Throughout the remainder of the paper, we shall consider the
frozen-matched choice as the best in the set of conformally flat,
longitudinal possibilities.

In Fig.\ \ref{fig:energyAa}, what is more important than the
comparison of prescriptions for hypersurface data is the fact that
none of them is very good.  Figure \ref{fig:energyAb} gives some
insight into the details of the failure. Here, for infall from
$r_0=1.5(2M)$, the data for $\psi$ on each hypersurface, are replaced
by frozen data; the true data for $\dot{\psi}$ (i.e., the data
evolved from $t=0$) are retained.  It is important and interesting that
the replacement of the $\psi$ data introduces a much smaller error than
the replacement of $\dot{\psi}$. This is a similar conclusion that one
finds in the results of Baker {\em et al.}\cite{bakeretal} in a rather
different context.

The previous energy results, of course, are for infall from an
astrophysically unreasonable initial radius. In Fig.\ 
\ref{fig:energyBc}, energy results are shown for infall from
$r_0=15(2M)$, in which case the true total quadrupole radiated is
$1.64\times10^{-2}m_0^2/(2M)$. Results are shown for three choices of
prescribed hypersurface data. Again we see that the agreement with the
true energy is reasonably good if the true (i.e., evolved) data for
$\dot{\psi}$ is retained. 
For the other cases the agreement is good
only if the data is replaced on a hypersurface well before the particle
reaches the peak of the potential $V_\ell$ at around $1.55(2M)$.
 
We ask next what the appearance of hypersurface data is, and how the
true and prescribed forms differ.  In Fig.\ \ref{fig:psiAa} we show
the form of the true quadrupole $\psi$ generated by a particle
falling from rest at $r_0=15(2M)$, with Misner data initially specified.
The figure shows the manner in which $\psi(r^*)$ evolves from its initial
Misner form at $t=0$. As the particle moves inward, the simple
prescribed form of a single peak evolves into more complex shapes.
Figure \ref{fig:psiAb} shows that on late hypersurfaces, after the
particle, at $t\approx100, r^*_p\approx0$ has passed through the
region in which the Zerilli potential is strong, $\psi$ clearly
contains the shape of in- and outgoing waves with the profile of
quasinormal ringing.

Figures \ref{fig:psiBa} -- \ref{fig:psiBc} compare the true $\psi$ on
a hypersurface with prescribed horizon frozen data on the same
hypersurface. It should be noted that the magnitude of the
discontinuity in the $r$ derivative of $\psi$ is related, through the
hamiltonian constraint, to the mass of the particle. It must,
therefore, be the same for the pairs of curves in these figures, and
for all forms of $\psi$ satisfying the hamiltonian constraint. Results
are shown for several different hypersurfaces, and the meaning is
clear. The prescribed $\psi$ never contains the complexity of shape
that indicates the presence of radiation.

In Fig.\ \ref{fig:psiCa}, $\psi$ is shown for Misner prescribed data on
a sequence of hypersurfaces, this time for infall from $r_0=1.5(2M)$.
In Fig.\ \ref{fig:psiCb} the frozen prescribed $\psi$ solution is
shown on the same hypersurfaces. The comparison of the two figures
shows the difference induced by the freezing of the value of $\psi$ at
the horizon.

The differences in the evolved and prescribed forms of $\psi$ suggest
that at late times the form of $\psi$ is not conformally flat initial
data. This is explicitly demonstrated and quantified in Figs.\ 
\ref{fig:nonconfa} and \ref{fig:nonconfb}, in which the gauge invariant
index of conformality, developed in Sec.\ II.B is plotted.  The
traveling bumps, with the appearance of radiation, confirm that the
wave content of the true initial data is associated with its failure
to be conformally flat.

We next look at results for $\dot{\psi}$. Figure\ \ref{fig:psidotAa},
\ \ref{fig:psidotAb},\ \ref{fig:psidotAc} compare the true
(i.e., evolved) form of the
quadrupolar $\dot{\psi}$ with the horizon matched prescription, for
infall from rest (and Misner data) at $t=0, r_0/2M=15$. As would be
expected, the difference between the evolved and the prescribed forms
grows with time.

The same comparison of true and horizon-matched $\dot{\psi}$ is given
in Fig.\ \ref{fig:psidotBa} and \ref{fig:psidotBb}, this time for
infall from $r_0/2M=1.5(2M)$.
In all the comparisons of $\dot{\psi}$, the horizon matched choice was
made for the prescribed data. The specific choice, however, makes
little difference in the comparisons.  In Fig.\ \ref{fig:psidotC} a
comparison of the true $\dot{\psi}$ and two prescriptions is given for
infall from $r_0/2M=1.5$. (Infall from the small radius magnifies
differences of evolution and prescription.) The difference between
the basic Bowen-York\cite{bowenyork} prescription for $\dot{\psi}$, and
the horizon matched prescription, is small, while the difference
between either of them and the correct data for $\dot{\psi}$ is very
large. The horizon-matched prescription is a slightly better approximation
than the ``basic'' one.

Another way of looking at the difference between the true data and
prescribed data is to investigate the effect the difference has on
outgoing radiation. Figures Fig.\ \ref{fig:waveformAa},
\ \ref{fig:waveformAb},\ 
\ref{fig:waveformAc} show the waveforms (that is, $\psi(t)$ at large
$r$) generated by different types of prescribed data. In each case, the
solid curve, labeled $t=0.0$, shows the ``true'' waveform, the
waveform generated by evolution of the original, momentarily
stationary data at $t=0$. This true waveform is compared with the
waveforms evolved from prescribed data placed on later hypersurfaces.
The figures show that prescribed data on the $t/2M=63.9$ hypersurface
leads to an outgoing waveform in reasonably good agreement with the
true waveform. For prescribed data on $t/2M=93.6$ (with the particle
at $r_p^*/2M=3.91$, or $r_p/2M=3.15$), however, large differences are
evident between waveform and the true waveform. The disagreement is
less severe when the evolved data for $\dot{\psi}$ is retained.

An analogous comparison of waveforms is shown in Fig.\ 
\ref{fig:waveformB} for infall from $r_0/2M=1.5$. Here the effect of
prescribed data on the phase of the wave is much more evident than in
the $r_0/2M=15$ case.

In the work of Abrahams and Cook\cite{abcook}, one of the motivations
for the present work, the ``close limit'' was used to evolve
prescribed data.  This technique\cite{pp,ap1} is applicable to a
hypersurface at a time late enough that the colliding bodies can be
considered to be inside a single nearly spherical horizon. In this
technique one uses only the large-$r$ parts of the initial data of the
colliding bodies. This technique was found to be surprisingly
successful in dealing with head on collisions of equal mass holes. We
saw in Paper I, however, that for very unequal mass holes the close
limit is valid only at extremely small separations. For comparison, we
have applied the close limit to the present problem. On a sequence of
hypersurfaces we replace the hypersurface data by the large-$r$ form
of the data, and evolve the results. The resulting energies are shown
in Fig.\ \ref{fig:close}.  These results, like those in Paper I, show
that the close-limit calculations of energy are of limited usefulness.

\section{Discussion and conclusions}

The numerical results of the previous section give a range of
comparisons between astrophysically evolved, and ``prescribed''
hypersurface data (for quadrupole modes.)  The first lesson to be
learned from these results is that for the particle limit the standard
prescriptions for hypersurface data are not adequate for describing an
astrophysical strong gravitational field. For a particle falling from
$r_0=15(2M)$, for example, an approximation for the radiation using
prescribed data on a late hypersurface seriously overestimates the
radiated energy.  Figure \ref{fig:energyBc} shows that prescribed data
for the particle at $r_p=3.6M$ (equivalent to $r_p^*/2M=1.6$) gives
a radiated energy too large by a factor of 10. Even for a hypersurface
with the particle in the marginally strong field at $r_p=5.7M$
(equivalent to $r_p^*/2M=3.4$) the energy is overestimated by a
factor of two.

This is in marked contrast to the results of Abrahams and
Cook\cite{abcook} for the head on collision of equal mass holes.  They
prescribed standard data on a late time hypersurface and found
predictions of energy in excellent agreement with predictions from
numerical relativity.  Baker and Li\cite{bakerli}, looking further
into this problem, found that the predicted energy was remarkably
insensitive to the choice of hypersurface. Good predictions could be
made with data specified on hypersurfaces over a fairly wide range of
times. There are certain technical differences between the nature of
the prescribed data we use here, and that used by Abrahams and Cook.
In particular, they used fully symmetrized data, whereas our preferred
choice of prescribed data is data that is conformally flat,
longitudinal data that is frozen and matched at the horizon. This is
surely not the origin of the different conclusions. For one thing
switching from symmetrized hypersurface data to antisymmetrized, in
numerical relativity computations, makes only a minor difference in
the results.  More important, the work of Baker and Li does not use
symmetrized data, and the study by Baker {\em et al.}\cite{bakeretal}
finds excellent results when symmetrized data is replaced by
unsymmetrized approximate data. As a further check of this point, we
have redone several of our computations using symmetrized prescribed
data and, as expected, found only minor differences from the results
for other prescriptions.

The crucial difference between the failure of prescribed data here,
and its success in the previous studies, must lie in the difference
between the head on collision of equal mass throats, and of a large
mass throat with a small mass (``particle'') one. It is not difficult
to see why this should be. It is the same reason the ``close limit''
approximation is not successful in the collision of very unequal
masses. (See Fig.\ \ref{fig:close} and Paper I.)
For the collision of equal masses, a common
horizon engulfs both colliding throats as they begin to get into each
other's strong field influence. The generation of the large amplitude
gravitational waves (i.e., the excitation of quasinormal ringing)
more-or-less coincides with the disappearance of the individual
throats inside a single connected horizon. For a hypersurface
corresponding to both throats inside a single horizon, it is plausible
(and is found to be valid by the success of the calculations) that
details of the geometry near the throats is not important since they
cannot influence the region exterior to the horizon. For a collision
of high and low mass throats the situation is very different. Strong
radiation is generated when the particle is near a coordinate distance
$r=1.5(2M)$, near the peak of the Zerilli potential. For our
$t=$\,constant slicing this corresponds to a hypersurface for which
there is no common horizon, i.e., the particle is still well outside
the horizon of the background hole. For this slicing, in fact, the
particle {\em always} stays outside the background horizon, and the
details of its local field are always in causal connection with
infinity.  The fault would not seem to be in the slicing. A
$t=$\,constant slicing was the choice in the method used by Abrahams
and Cook\cite{abcook} and by Baker and Li\cite{bakerli}, and is an
obviously natural slicing for the particle infall problem.

We conclude from these considerations that the fault lies in the
nature of the prescribed data we have been using, i.e., the standard
choice of conformally flat longitudinal hypersurface data. As shown in
the figures of the previous section, there is a dramatic difference
between the true hypersurface data on a late hypersurface, and
prescribed data. There are two strong reasons to suspect that the most
important difference between true hypersurface data and any of the
prescribed data is the difference in $\dot{\psi}$. First, our argument
above suggests that the failure of prescribed data in the particle
limit lies in the failure to describe fields near the particle. In
Figs. \ref{fig:psidotAb} -- \ref{fig:psidotC}, we see that there is
(at least visually) a large difference between the true and the
prescribed data close to the particle. This difference can be ascribed
to the constraints on conformally flat longitudinal data.  The sharp
variations in $\dot{\psi}(r^*)$ are, intuitively, features that should
couple strongly to radiation. The inability of the prescribed data to
model these features, must be viewed as potentially important. The
second reason for focusing attention on $\dot{\psi}$ is that the
hypersurface data for $\dot{\psi}$ appears to be much more important
than that for $\psi$. This was the conclusion in the study by Baker
{\em et al.}\cite{bakeretal}. In that work, which used colliding
throats, and no stress energy, Einstein's equations were linearized in
the momentum of the throats and a clear identification could be made
of how much of the radiated energy could be ascribed to the initial
3-geometry, how much to the momentum, i.e., the initial extrinsic
curvature, and how much to the interaction of the two.  It was found
that except for extremely small initial momentum the radiation was
almost completely due to the extrinsic curvature. This was called
``momentum dominance'' by Baker {\em et al.}  Here we cannot make such
a clear distinction. Due to the moving particle source, treated as a
stress energy contribution, we cannot ascribe the radiation to the two
kinds of initial data information. Nevertheless, the energy results in
our Figs.\ \ref{fig:energyAa} -- \ref{fig:energyBc} show that some
other form of ``momentum dominance'' applies to the particle infall;
the energy radiated is much more sensitive to the details of the
extrinsic curvature than to the details of the 3-geometry.

The insights provided by the particle limit have helped to clarify
what is needed in the way of hypersurface data for numerical
evolution.  A direct consequence of the present study is the
realization that prescribed data will not work as successfully for
unequal mass head on collisions as for the equal mass case. At some
ratio of masses of the colliding holes the use of prescribed data on a
late time hypersurface will start to give a significant overestimate
of the radiated energy. We have, so far, studied only the quadrupole
perturbations. Higher multipole moments are less important
astrophysically, and are not likely to lead to very different
conclusions.

The head on collision, of course, is neither astrophysically
plausible, nor an interesting source of outgoing radiation. The
interesting case is the last stage in the orbital decay of a binary
pair of holes. For this one would like to start with astrophysically
reasonable data on a ``as late as possible'' hypersurface. An
understanding of how to do this for the head on case is a necessary
step (and perhaps a sufficient one) towards an understanding of the
more general problem. With that motivation, we shall, in a subsequent
paper, investigate what {\em can} be done to provide good late time
data. We shall, in particular, abandon the traditional choice of
conformally flat, longitudinal data, and, instead, shall focus on data
that gives a good description of the field near the particle.

\begin{figure}[h]
\caption{Geometry in the conformally related flat space.}\label{fig:geomforK}
\end{figure}

\begin{figure}[h]
\caption{A cell of the computational grid containing a segment
  of the particle worldline. Grid nodes are shown coordinatized with
  $t,r^*$ labels, and with values of the null coordinates $u\equiv
  t-r^*$ and $v\equiv t+r^*$. The areas $A_1\cdots A_4$ are used as
  weights in the numerical algorithm.}\label{fig:rombo}
\end{figure}

\begin{figure}[h]
\caption{
  Hypersurfaces of constant $t$ in the Schwarzschild spacetime.  The
  trajectory is shown of a particle starting from rest at position
  $r_0/2M=15$, and falling inward on a radial geodesic. Also shown are
  hypersurfaces of constant time at $t/2M=20, 40, 60$ and 80. The
  hypersurfaces are labeled with the value of $t$, of the particle
  position $r_p$, and of momentum parameter $P$.  The $t/2M=20$
  hypersurface, for example, can also be referred to as the $P/m_0=0.05$
  hypersurface, the $r_p/2M=14.6$ hypersurface, or the $r_p^*/2M=17.2$
  hypersurface.
  }\label{fig:traj}
\end{figure}

\begin{figure}[h]
\caption{
  Radiated Energy, for infall from $r_0=1.5(2M)$ Energy is shown as a
  function of the $r_p^*$ label of a hypersurface and results are
  given for several types of prescribed hypersurface data.  Radiated
  energy computed is always the energy radiated starting from the
  time of the hypersurface on which data is prescribed. See text for
  details.  }\label{fig:energyAa}
\end{figure}

\begin{figure}[h]
\caption{
  Radiated Energy, for infall from $r_0=1.5(2M)$. Energy is shown as a
  function of the $r_p^*$ label of a hypersurface. On that
  hypersurface the true, i.e., evolved, data for $\dot{\psi}$ has been
  retained, but frozen prescribed data for $\psi$ has been substituted
  for the true $\psi$.}
\label{fig:energyAb}
\end{figure}

\begin{figure}[h]
\caption{
  Radiated Energy, for infall from $r_0=15(2M)$. Energy is shown as a
  function of the $r_p^*$ label of a hypersurface and results are
  given for several types of prescribed hypersurface data. The computed
  energy is always the energy radiated starting from the
  time of the hypersurface on which data is prescribed. See text for
  details.  }\label{fig:energyBc}
\end{figure}

\begin{figure}[h]
\caption{The true quadrupole $\psi(r^*)$ for infall from $r_0=15(2M)$, 
  i.e., $\psi$ computed by numerical evolution from the initial
  hypersurface. Curves are shown for several hypersurfaces labeled
  with $r^*_p$, the value of the particle's $r^*$ location on each
  hypersurface.  }\label{fig:psiAa}
\end{figure}
\begin{figure}[h]
\caption{The true quadrupole $\psi(r^*)$ for infall from $r_0=15(2M)$, 
shown on late hypersurfaces. For $t/2M=109.7$ the shape of $\psi$ shows
outgoing quasinormal radiation; for $t/2M=149.6$ both ingoing and outgoing
quasinormal oscillations are evident.}\label{fig:psiAb}
\end{figure}

\begin{figure}[h]
\caption{
  The comparison of true and prescribed quadrupole $\psi$ for
  infall from $r_0=15(2M)$. The form of the true data $\psi(r^*)$ for
  infall from Misner data on the original hypersurface, is compared with the
  frozen prescribed data on the hypersurface at $t/2M=52.2$.
  }\label{fig:psiBa}
\end{figure}
\begin{figure}[h]
\caption{For infall from $r_0=15(2M)$, the evolved data is compared
with the frozen prescribed data, at $t/2M=96.9$}\label{fig:psiBb}
\end{figure}
\begin{figure}[h]
\caption{For infall from $r_0=15(2M)$, the evolved data is compared
with the frozen prescribed data, at $t/2M=100.2$.}
\label{fig:psiBc}
\end{figure}

\begin{figure}[h]
\caption{Misner prescribed data for infall from $r_0=1.5(2M)$, shown
on a sequence of hypersurfaces.}
\label{fig:psiCa}
\end{figure}
\begin{figure}[h]
\caption{Frozen prescribed data for infall from $r_0=1.5(2M)$, shown
on a sequence of hypersurfaces.}
\label{fig:psiCb}
\end{figure}

\begin{figure}[h]
\caption{The index of conformality $I_{\rm conf}$ for infall from
  $r_0=15(2M)$ with Misner initial data. The index, for
  quadrupole perturbations, is equivalent to $H_2-K$ in the RW gauge,
  and is a gauge invariant measure of the extent to which the
  hypersurface 3-geometry fails to be conformally flat (see text). The
  index is given, as a function of $r^*$, for several different
  hypersurfaces. The noisy nature of the curves is caused by the need
  to take second differences of numerical results to compute $I_{\rm
    conf}$.}\label{fig:nonconfa}
\end{figure}
\begin{figure}[h]
\caption{The index of conformality $I_{\rm conf}$ for infall from
 $r_0=15(2M)$, shown on late hypersurfaces.}
\label{fig:nonconfb}
\end{figure}

\begin{figure}[h]
\caption{For $r_0/2M=15$, the true (i.e., evolved) form of $\dot{\psi}$
compared with the
  horizon matched prescribed form of $\dot{\psi}$, for
  $t/2M=52.2$.}
\label{fig:psidotAa}
\end{figure}
\begin{figure}[h]
\caption{For $r_0/2M=15$, the  true (i.e., evolved) form of $\dot{\psi}$
compared with the
  horizon matched prescribed form of $\dot{\psi}$, for
  $t/2M=96.9$.}
\label{fig:psidotAb}
\end{figure}
\begin{figure}[h]
\caption{For $r_0/2M=15$, the true (i.e., evolved) form of $\dot{\psi}$
compared with the
  horizon matched prescribed form of $\dot{\psi}$, for
  $t/2M=100.2$.}
\label{fig:psidotAc}
\end{figure}

\begin{figure}[h]
\caption{For $r_0/2M=1.5$, the true (i.e., evolved) form of $\dot{\psi}$
compared with the
  horizon matched prescribed form of $\dot{\psi}$, for
  $r_p/2M=1.34$.}
\label{fig:psidotBa}
\end{figure}
\begin{figure}[h]
\caption{For $r_0/2M=1.5$, the  true (i.e., evolved) form of $\dot{\psi}$
compared with the
  horizon matched prescribed form of $\dot{\psi}$, for
  $r_p/2M=1.21$.}
\label{fig:psidotBb}
\end{figure}

\begin{figure}[h]
\caption{For $r_0/2M=1.5$, the  true, i.e., evolved, form of $\dot{\psi}$
compared with the
  horizon matched prescribed form of $\dot{\psi}$, and with the
  Bowen-York form.}
\label{fig:psidotC}
\end{figure}

\begin{figure}[h]
\caption{The quadrupole waveform, $\psi(t)$ at $r/2M=1000$, for 
  infall from $r_0/2M=15$. Three curves are shown corresponding to the
  true waveform (data prescribed at $t/2M=0$) and to the waveforms
  generated when frozen $\psi$ data and horizon matched $\dot{\psi}$
  data are imposed on later hypersurfaces.}
\label{fig:waveformAa}
\end{figure}
\begin{figure}[h]
\caption{The quadrupole waveform, $\psi(t)$ at $r/2M=1000$, for 
  infall from $r_0/2M=15$. Three curves are shown corresponding to the
  true waveform (data prescribed at $t/2M=0$) and to the waveforms
  generated when horizon matched $\dot{\psi}$ data are imposed on
  later hypersurfaces and the true, i.e., evolved, values of $\psi$
  are retained.}
\label{fig:waveformAb}
\end{figure}
\begin{figure}[h]
\caption{The quadrupole waveform, $\psi(t)$ at $r/2M=1000$, for 
  infall from $r_0/2M=15$. Three curves are shown corresponding to the
  true waveform (prescribed data a $t/2M=0$) and to the waveforms
  generated when frozen $\psi$ data are imposed on later hypersurfaces
  and the true, i.e., evolved, values of $\dot{\psi}$ are retained.}
\label{fig:waveformAc}
\end{figure}

\begin{figure}[h]
\caption{The quadrupole waveform, $\psi(t)$ at $r/2M=1000$, for 
  infall from $r_0/2M=1.5$. The solid curve shows the ``true''
  waveform evolved directly from the initial data at $t/2M=0$. The
  dashed curve gives the waveform generated when prescribed data
  (horizon-frozen data) for $\psi$, and evolved data for $\dot{\psi}$ are
  placed on a late hypersurface.}
\label{fig:waveformB}
\end{figure}

\begin{figure}[h]
\caption{Radiated energy for infall from $r_0=15(2M)$ computed with 
the close limit. On every hypersurface close limit data was substituted
and evolved. The computed energy is plotted as a function of the hypersurface
parameter on which the close limit data was substituted.}
\label{fig:close}
\end{figure}

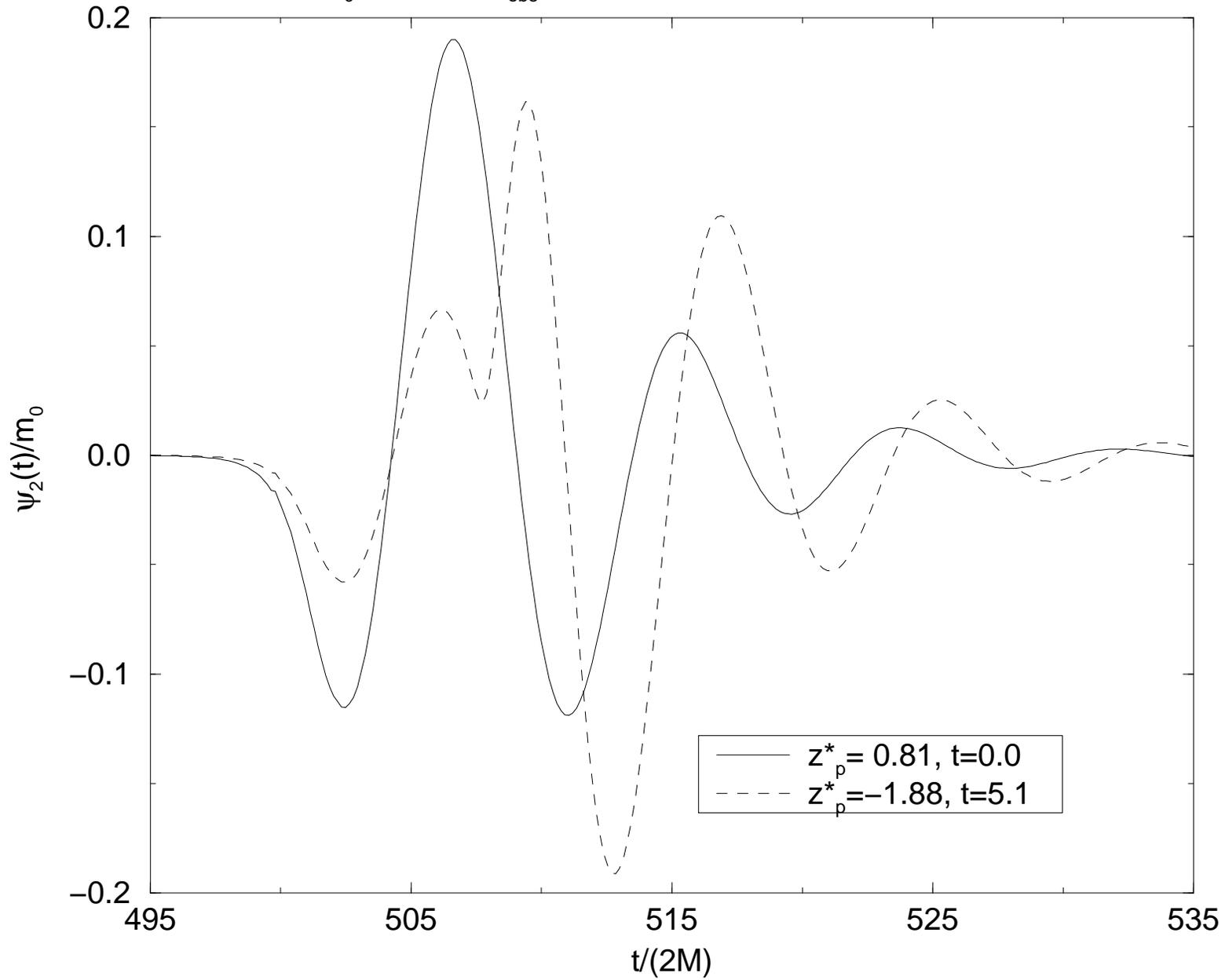
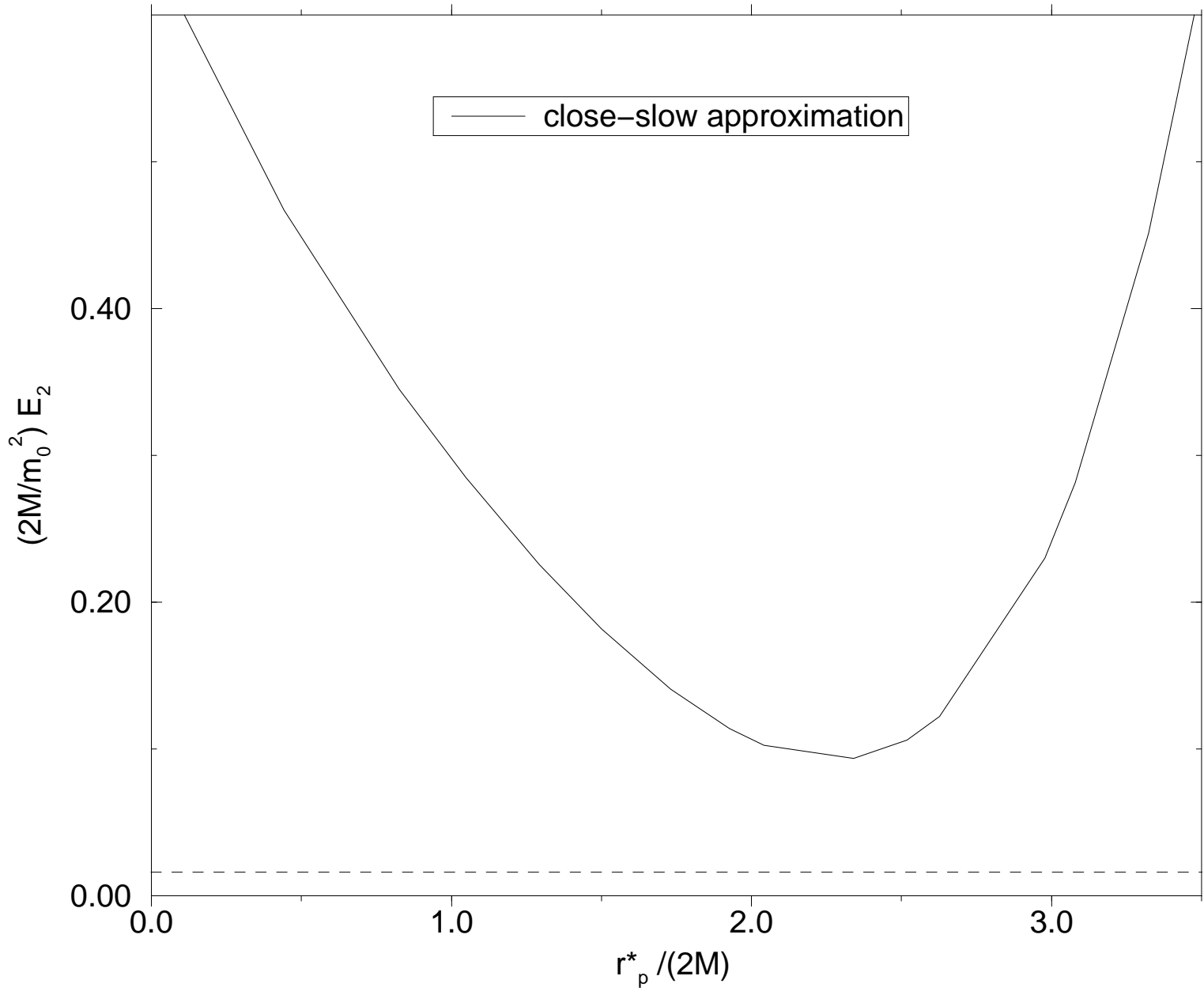
\end{document}